\documentclass[aps,amsmath,amssymb,prd,showpacs,floatfix,preprint,superscriptaddress,nofootinbib,12pt]{article}
\usepackage{jheppub}
\usepackage{verbatim}

\usepackage{graphicx}
\usepackage{subfigure}
\input{epsf}
\usepackage{epsfig}
\usepackage{epstopdf}
\usepackage{amsthm}
\usepackage{slashed}

\def\({\left(} \def\){\right)}
\def\[{\left[} \def\]{\right]}
 
\def\del{{\partial}}

\newcommand{\non}{\nonumber \\}

\newcommand{\be}{\begin{equation}}
\newcommand{\ee}{\end{equation}}
\newcommand{\bea}{\begin{eqnarray}}
\newcommand{\eea}{\end{eqnarray}}
\newcommand{\ba}{\begin{eqnarray}}
\newcommand{\ea}{\end{eqnarray}}

\newcommand{\beq}{\begin{equation}}
\newcommand{\eeq}{\end{equation}}
\newcommand{\beqa}{\begin{eqnarray}}
\newcommand{\eeqa}{\end{eqnarray}}
\newcommand{\beqar}{\begin{eqnarray*}}
\newcommand{\eeqar}{\end{eqnarray*}}

\newcommand{\reef}[1]{(\ref{#1})}

\newcommand{\eg}{{\it e.g.,}\ }
\newcommand{\ie}{{\it i.e.,}\ }

\newcommand{\mt}[1]{\textrm{\tiny #1}}

\title{On fast quenches and spinning correlators}

\author{Mikhail Goykhman}
\emailAdd{michael.goykhman@mail.huji.ac.il}
\author{Tom Shachar}
\emailAdd{tom.shachar@mail.huji.ac.il}

\author{and Michael Smolkin}
\emailAdd{michael.smolkin@mail.huji.ac.il}
\affiliation{The Racah Institute of Physics, The Hebrew University of Jerusalem, \\ Jerusalem 91904, Israel}

\abstract{ We study global quantum quenches in a continuous field theoretic system with UV fixed point. Assuming that the characteristic inverse time scale of the smooth quench is much larger than all scales inherent to the system except for the UV-cutoff, we derive the universal scaling behavior of the two-point correlation functions associated with Dirac fields and spin-1 currents. We argue that in certain regimes our results can be recovered using the technique of operator product expansion.
}

\begin{document}
\maketitle

\section{Introduction}

Quantum quench is a unitary process during which a physical system,
typically prepared in the ground state of the unperturbed Hamiltonian, is subject to an evolution under a prescribed time-dependent change in the parameters of the Hamiltonian. Thus, for instance, one can think of varying the couplings inherent to the system or introducing a time-dependent background field into the Hamiltonian. The specific choice of the time-dependent profile for these parameters is conventionally referred to as a quench protocol, and the so-called quench rate is used to classify various scenarios. Usually the quench rate is identified with a characteristic inverse time scale, $\delta t^{-1}$, over which the parameters experience a significant change. 

The aspiration to understand the non-equilibrium dynamics in general and mechanism of relaxation in particular is one of the major theoretical motivations behind the studies of quantum quenches. Thus, for instance, one of the particularly interesting class of quenches drives the system through a critical phase \cite{Kibble:1976sj,Zurek:1985qw}, where the dynamics is governed by a conformal field theory (CFT). When the system is sufficiently close to criticality the quench rate $\delta t^{-1}$ becomes large compared to any other scale in the system, and therefore adiabatic approximation breaks down. As a result, the entire system
is driven far away from equilibrium, and its subsequent relaxation is in the spot light of both experimental and theoretical research. 

Moreover, the interest in quantum quenches has been recently increased due to the successful 
experiments with cold atoms trapped in optical lattices
\cite{Greiner:2002,Bloch:2007,Polkovnikov:2010,Cazalilla:2011,Mitra:2017}.
Such systems exhibit a quantum critical regime, and
 can be driven through a critical point by changing the optical lattice spacing,
while preserving the quantum coherence of the system for a sufficiently long time.
Therefore these systems serve as an ideal experimental setup for the study of quantum
quenches.

Remarkably, quantum quenches reveal a unique laboratory where the dynamics of thermalization can be studied. Of course, if the initial state of the system is pure, it remains pure at all times due to the unitary evolution. However, this is not true if the state is reduced to a small subsystem, or the system as a whole respects the Eigenstate Thermalization Hypothesis \cite{Srednicki94}. In these cases it is tempting to address the question whether equilibration process bears universal merits, and estimate, for example, the characteristic time it takes for the system to approach the equilibrium state described by a certain thermal ensemble \cite{Polkovnikov:2010,Cazalilla:2011,Fagotti2013,Mitra:2017}.

In fact, the observables, such as the vacuum expectation values and correlation functions of
the physical operators in the quenched two-dimensional quantum field theories, have been extensively studied
in the literature. One of the earlier works in this direction considers a
system which is prepared in the ground state of the Hamiltonian
$H_\lambda=H_\mt{CFT}+\lambda{\cal O}$,
where ${\cal O}$ is a relevant scalar operator
and $H_\mt{CFT}$ governs the dynamics of a CFT
\cite{Calabrese:2006rx,Calabrese:2007rg}.
At $t=0$ the coupling $\lambda$ is instantaneously tuned to zero,
and the subsequent relaxation of the system is studied.
It has been demonstrated that relaxation of the observables following the
instantaneous (also known as `sudden') quantum quench, exhibits a universal behavior
governed by the CFT scaling dimensions \cite{Calabrese:2006rx,Calabrese:2007rg}; also
see \cite{Delfino:2014qfa,Delfino:2016bln} for recent developments in perturbative formulation of the instantaneous quan-
tum quench problem near criticality in the 1 + 1-dimensional case. 

The opposite regime of smooth rather than sudden quenches is not tractable in general. Holography, however, provides a necessary toolkit where the quench dynamics with a finite quench rate can be addressed. Thus, for instance, inspired by the earlier works of \cite{Chesler:2008hg,Basu:2011ft}, the authors of \cite{Buchel:2012gw,Buchel:2013lla} used  numerical methods in the holographic setup to study the response of a strongly-coupled CFT to a smooth quantum quench of the scalar and fermionic mass. The dimensionless parameter $T\delta t$, where $T$ is the temperature of the initial state, was used by the authors to distinguish between the fast ($T\delta t\ll 1$) and slow ($T\delta t\gg 1$) quenches. In the case of a fast quench it has been found that the observables in the system, such as the one-point correlation function $\langle\mathcal{O}\rangle$ of an operator adjoint to the quenched parameter $\lambda$,
exhibit a new universal scaling behavior with respect to the quench rate. This conclusion has been further generalized analytically in \cite{Buchel:2013gba}, concluding that the fast quench of a strongly-coupled CFT in $d$ dimensions manifests a universal scaling at early times, \eg $\langle\mathcal{O}\rangle\sim \lambda\delta t^{d-2\Delta}$, where ${d/2}<\Delta < d$ is the conformal dimension of the scalar operator $\mathcal{O}$.

The universal scaling behavior 
has further been shown to exist in the free quantum field theories \cite{Das:2014jna,Das:2014hqa,Das:2015jka},
where response of the system to the quench of a mass has been studied.
It has subsequently been argued that the universal scaling is a general property inherent to any
quantum field theory following the quench dynamics \cite{Das:2016lla,Das:2016eao,Das:2017sgp,Dymarsky:2017awt}.
In other words, the response of an operator to the quantum quench is determined by the
ultra-violet (UV) CFT properties of the system, namely the UV conformal dimension of that operator.
To extend and generalize the results of \cite{Das:2016lla,Das:2016eao,Das:2017sgp}, the authors of \cite{Dymarsky:2017awt} scrutinized the response of one- and two-point correlation functions of scalar operators in the framework
of conformal perturbation theory around a generic CFT.

One salient feature of the universal scaling law $\langle {\cal O}\rangle \sim \lambda\delta t^{d-2\Delta}$
exhibited by the systems subjected to a fast quantum quench, is its singular behavior
in the instantaneous quench limit $\delta t\rightarrow 0$, for the operators of scaling dimension
$\Delta\in \left(\frac{d}{2},d\right)$. This is contrasted with the finite behavior in the case of an instantaneous
quench \cite{Calabrese:2006rx,Calabrese:2007rg}. Such a discrepancy has been
argued to follow from the non-commutativity of the instantaneous quench limit, $\delta t\rightarrow 0$,
and the limit of taking the UV cutoff to infinity \cite{Das:2016lla,Das:2017sgp}.

%\vspace{0.5cm}

In this paper we study the effect of quantum quenches on the correlation functions of spin-1 and spin-1/2 operators in a theory with UV fixed point. We assume that the quench rate is the shortest scale compared to any other scale inherent to the system (except for the UV cutoff). As argued in \cite{Dymarsky:2017awt}, in this regime the quenched correlation functions at early times are dominated by the vicinity of the UV fixed point provided that the typical distance between the operators is sufficiently small. In particular, one can employ the conformal perturbation theory to study the effect of quench on the correlators.  Following this approach we derive the universal scaling behavior of the spinning correlation functions in various regimes.

The rest of the paper is organized as follows. In section \ref{sec:conf_pert_theory} we briefly review the essentials
of the perturbation theory used by us in the context of quantum quenches. In section \ref{section:currents} we study the linear response of the quenched current-current correlation functions. The scaling dimensions of the currents are arbitrary, and therefore they are not necessarily conserved.  In the limit of fast but smooth quenches we find that correlation functions scale universally with $\delta t$. We point out that in certain regimes our results can be derived using the OPE techniques.  In section \ref{sec:fermions} we repeat a similar set of calculations for the correlation functions of two spinors and find qualitative similarity between the results obtained for the currents in section \ref{section:currents} and for the scalars in \cite{Dymarsky:2017awt}.

\section{Preliminary remarks}
\label{sec:conf_pert_theory}

In this section we outline the quench protocol and briefly overview the necessary formalism of conformal perturbation
theory that will be used in the next sections. 

Consider a $d$-dimensional QFT deformed by the scalar operator ${\cal O}$ 
\begin{equation}
\label{quenched_hamiltonain}
H=H_0+\lambda(t)\,\int d^{d-1}{\bf x}\,{\cal O}({\bf x})\,,
\end{equation}
where $H_0$ denotes the Hamiltonian of the unperturbed QFT, whereas the quench protocol has the form
\begin{equation}
\lambda(t)=\delta \lambda \,f(\xi)\,,\qquad \xi=\frac{t}{\delta t}\,,\quad \delta \lambda\sim \ell^{\Delta-d}\, ,
\label{quench_profile}
\end{equation}
where $\Delta$ is the scaling dimension of ${\cal O}$, $f(\xi)$ is a smooth pulse function supported on the interval $\xi\in (-1,1)$ and $\ell$ is a characteristic length scale introduced by the quench into the state of the QFT. This profile represents a quantum bump of characteristic width $\delta t$.

We assume that initially the system resides in the vacuum state $|0\rangle$ of the QFT governed by $H_0$
\begin{equation}
|\Psi(t)\rangle \underset{t\to-\infty}{\longrightarrow}|0\rangle\,.
\label{intcond}
\end{equation}
Of course, in the absence of external deformation the system clings to the vacuum state forever. However, the quench typically results in a complicated dynamics.  Expanding the state of the system in power series in $\lambda(t)$, yields 
\begin{equation}
\label{psi_t}
|\Psi (t)\rangle =e^{-iH_0(t-t')}\left(1 -i\,\int _{t'}^tdt_1\,\lambda(t_1){\cal O}(t_1) + \ldots\right)\,|\Psi(t')\rangle\,,
\end{equation}
where ${\cal O}(t)$ represents the Heisenberg operator 
\begin{equation}
\label{heisenberg_operators}
{\cal O}(t)=\int d^{d-1}{\bf x}\,{\cal }O(t,{\bf x})\,,
\quad {\cal O}(t,{\bf x})=e^{iH_0(t-t')}{\cal O}({\bf x})e^{-iH_0(t-t')}\,.
\end{equation}

The above expansion is formal and needs justification. In fact, it cannot be truncated in general. However,  in sections \ref{section:currents} and \ref{sec:fermions} we are going to use
\reef{psi_t} to calculate the linear response of the spinning correlators under the assumption that the quenched QFT has an UV fixed point, and $\delta t$ is the shortest scale in the system (except for the UV cutoff) satisfying $\delta\lambda\delta t^{d-\Delta}\ll 1$, where $d/2<\Delta<d$ is the scaling dimension of $\mathcal{O}$ at the UV fixed point. In this case, as argued in \cite{Dymarsky:2017awt} (see also earlier works \cite{Buchel:2013gba,Das:2016eao}), the correlation functions are dominated by the UV CFT, and the leading order effect can be derived by replacing $|0\rangle$ and $H_0$ with conformal vacuum and conformal Hamiltonian, $H_\mt{CFT}$, respectively.

\section{Quenched currents}
\label{section:currents}

Let us consider a QFT governed by the Hamiltonian \reef{quenched_hamiltonain}. Motivated by the earlier works \cite{Buchel:2013gba,Das:2014jna,Dymarsky:2017awt} we aim at deriving the universal scaling of the correlation function of two not necessarily identical or conserved currents
\begin{align}
\label{quenched_current_current_expansion}
G_{\mu\nu}^{(JJ)}(t_{1},{\bf x}_{1};t_{2},{\bf x}_{2})\equiv\left\langle J^{(1)}_{\mu}(t_1,{\bf x}_1)J^{(2)}_{\nu}(t_2,{\bf x}_2)\right\rangle\,.
\end{align}
Both currents are associated with the unperturbed QFT governed by $H_0$ and the expectation value is taken in the state satisfying  \reef{intcond}, \reef{psi_t}. We assume that $H_0$ has conformal UV fixed point and the quench rate, $\delta t^{-1}$, is much larger than any other scale in the system. 

The linear response of the above current-current correlator to a quench protocol outlined in the previous section is given by
\begin{align}
\label{quenched_current_current_correction}
\delta^{(1)}G_{\mu\nu}^{(JJ)}&(t_{1},{\bf x}_{1};t_{2},{\bf x}_{2})=i\int _{-\infty}^{t_2}dt'\,\lambda(t')\int d^{d-1}{\bf y}
\langle[{\cal O}(t',{\bf y}),J^{(1)}_\mu(t_1,{\bf x}_1)J^{(2)}_\nu (t_2,{\bf x}_2)]\rangle_0\notag\\
&+i\int _{t_2}^{t_1}dt'\,\lambda(t')\int d^{d-1}{\bf y}
\langle[{\cal O}(t',{\bf y}), J_\mu^{(1)}(t_1,{\bf x}_1)]J_\nu^{(2)} (t_2,{\bf x}_2)\rangle_0\,.
\end{align}
where we combined \reef{intcond}, \reef{psi_t} with \reef{quenched_current_current_expansion} and the subscript $0$ indicates that the correlation functions are evaluated in the vacuum state of $H_0$.  This result can be also derived using the standard Keldysh-Schwinger path integral interpretation of \reef{quenched_current_current_expansion}.

As was argued in \cite{Buchel:2013gba,Das:2014jna,Dymarsky:2017awt}, at early times and rapid quench rate, \ie $\delta\lambda\delta t^{d-\Delta}\ll 1$, the full dynamics of the quenched QFT is dominated by the UV fixed point. In particular, as we lower the dimensionless parameter $\delta\lambda\delta t^{d-\Delta}$ ($\delta\lambda$ fixed while $\delta t\to 0$), the linear response function $\delta^{(1)}G_{\mu\nu}^{(JJ)}(t_{1},{\bf x}_{1};t_{2},{\bf x}_{2})$, with $|0\rangle$ and $H_0$ replaced by the conformal vacuum and $H_\mt{CFT}$ respectively, takes over the terms associated with either higher order corrections in $\delta\lambda$ or other scales inherent to the system. 

Hence, in the regime of fast and smooth quenches the response of the current-current two-point function is completely universal. It is determined by the linear term in $\delta\lambda$ which is dominated by the correlation function entirely fixed by the conformal symmetry
\begin{equation}
\label{vector_vector_scalar_definition}
G^{(JJ{\cal O})}_{\mu\nu}(x_1,x_2,x_3)=\left\langle J^{(1)}_\mu(x_1)J^{(2)}_\nu(x_2){\cal O}(x_3)\right\rangle_\mt{CFT}\,.
\end{equation}
The embedding space formalism \cite{Weinberg:2010fx,Costa:2011mg} is the most efficient way to calculate the above correlator. We delegate the details to Appendix \ref{appendix:jjO}. The final answer factorizes into a product of scalar factor and a scale-invariant tensor structure
\begin{align}
\label{current-current-scalar_general}
G^{(JJ{\cal O})}_{\mu\nu}(x_1,x_2,x_3)&=S^{(JJ{\cal O})}(x_1,x_2,x_3)\,T^{(JJ{\cal O})}_{\mu\nu}(x_1,x_2,x_3)\,,\\
S^{(JJ{\cal O})}(x_1,x_2,x_3)&=\frac{1}{x_{12}^{\Delta_{123}}\,
x_{13}^{\Delta_{132}}\,
x_{23}^{\Delta_{231}}}\,,
\label{scalar_pt}\\
T^{(JJ{\cal O})}_{\mu\nu}(x_1,x_2,x_3)&=c_1\(\eta_{\mu\nu}-\frac{2\,x_{12\,\mu}\,x_{12\,\nu}}{x_{12}^2}\)\nonumber \\
&+c_2\left(\frac{x_{13\,\mu}x_{12\,\nu}}{x_{13}^2}
-\frac{x_{12\,\mu}x_{23\,\nu}}{x_{23}^2}
-\frac{x_{12\,\mu}x_{12\,\nu}}{x_{12}^2}
+\frac{x_{12}^2}{x_{13}^2x_{23}^2}\,x_{13\,\mu}x_{23\,\nu}\right)\,,
\label{tensor_pt}
\end{align}
where $x_{ij}=x_i-x_j$ for $i,j=1,2,3$, $c_1$ and $c_2$ are constants, and for brevity we introduced the following notation
\begin{equation}
\Delta_{ijk}=\Delta_i+\Delta_j-\Delta_k\,,\quad i,j,k=1,2,3\,,
\end{equation}
where $\Delta_1$, $\Delta_2$ and $\Delta_3$ denote the scaling dimensions of the primary fields $J^{(1)}_\mu$, $J^{(2)}_\nu$ and $\mathcal{O}$ respectively. If one of the currents is conserved, say $J^{(1)}_\mu$, then the following constraints hold\footnote{The relation between $c_1$ and $c_2$ follows from $\Delta_1=d-1$ and
\be
0=\left\langle \partial^\mu J^{(1)}_\mu(x_1) \, J^{(2)}_\nu(x_2) {\cal O}(x_3)\right\rangle_\mt{CFT}=
 \frac{\big(c_2(\Delta_2-\Delta_3)-c_1\Delta_{132}\big)(x_{12\nu}x_{23}^2+x_{23\nu}x_{12}^2)}{(x_{12}^2)^{{\Delta_{123}\over 2}+1}\,(x_{13}^2)^{{\Delta_{132}\over 2}+1}\,(x_{23}^2)^{\Delta_{231}\over 2}}\,.
 \nonumber
\ee
See \cite{Costa:2011mg} for the analysis of conservation condition and conformal invariance in the case of general spin.
}
\be
 \Delta_1=d-1~,\quad c_2={\Delta_{132}\over \Delta_2-\Delta_3}\, c_1 ~.
 \label{cons_const}
\ee
If, however, both $J^{(1)}_\mu$ and $J^{(2)}_\mu$ are conserved then on top of the above constraints we also have $\Delta_2=d-1$.

Now let us evaluate the equal time correlation between the temporal components of the two currents. In this case \reef{quenched_current_current_correction} simplifies
\begin{equation}
\label{quenched_current_current_correction_equal_time}
\delta^{(1)}G_{00}^{(JJ)}(t,{\bf x};t,0)=i\,\int _{-\infty}^tdt'\,\lambda(t')\int d^{d-1}{\bf y}\,\left\langle \left[
{\cal O}(t',{\bf y}),J^{(1)}_{0}(t,{\bf x})J^{(2)}_{0}(t,0)\right]\right\rangle_\mt{CFT}\,.
\end{equation}
Note that the right ordering of operators within the three-point function on the right hand side is achieved by introducing a small imaginary component to the Lorentzian times. An operator that is to the `left' of another should have smaller imaginary part. In particular, the above expression can further be written as
\begin{align}
\label{J0J0time_separation}
\delta^{(1)}G_{00}^{(JJ)}(t,{\bf x};t,0)
&=-2\,{\rm Im}\,\int _{-\infty}^tdt'\,\lambda(t')\,\left(c_1\,\frac{J(t-t',{\bf x};\Delta_{132},\Delta_{231},d)}{|{\bf x}|^{\Delta_{123}}}\right.\\
&-\left.c_2\,\frac{(t-t')^2\,J(t-t',{\bf x};\Delta_{132}+2,\Delta_{231}+2,d)}{|{\bf x}|^{\Delta_{123}-2}}\right)\,,
\end{align}
where $J$ is defined and calculated in Appendix \ref{master}, see \reef{Japx} and \reef{J}. It has the following asymptotic behavior
\bea
J(t,{\bf x};\delta_1,\delta_2,d)\Big|_{\delta t/|{\bf x}|\ll1}&=&\pi^\frac{d-1}{2}\,
\frac{\Gamma\left(\frac{1-d+\delta _1}{2}\right)}
{\Gamma\left(\frac{\delta _1}{2}\right)}\,\frac{\delta t^{d-1-\delta_1}}{|{\bf x}|^{\delta_2}}\left(-(\xi-i\epsilon)^2\right)^\frac{d-1-\delta_1}{2}+(1\leftrightarrow 2) ~,
\non
\label{J_fast_limit}
\\
\label{J_integral_small_x}
J(t,{\bf x};\delta_1,\delta_2,d)\Big|_{\delta t/|{\bf x}|\gg1}&=&\frac{\pi^\frac{d-1}{2}\Gamma\left(\frac{\delta_1+\delta_2-d+1}{2}\right)
\delta t^{\frac{d-1-\delta_1-\delta_2}{2}}}
{\Gamma\left(\frac{\delta_1+\delta_2}{2}\right)(-(\,\xi-i\epsilon)^2)^{\frac{\delta_1+\delta_2-d+1}{2}}} ~ ,
\eea
where we used the dimensionless parameter $\xi=t/\delta t$. 

Note that the linear response function vanishes in the limit $\delta t\to 0$ if the time instant $ t \ll \ell$ is fixed. Therefore at late times one has to resort to higher orders in $\delta\lambda$. However, this is not true at early times. In this range the response function exhibits an interesting universal scaling behavior. Setting for simplicity $t=0$ and using (\ref{J_fast_limit}) and (\ref{J_integral_small_x}), we obtain\footnote{We rely on the identities
\begin{align}
%\label{taking_imaginary_part}
\lim_{\epsilon\rightarrow 0}(-\xi^2\pm i\epsilon)^p&=\xi^{2p}\,e^{\pm i\pi p}\,,\non
\Gamma(z)\Gamma(1-z)&=\frac{\pi}{\sin (\pi z)}\,.
\notag
\end{align}
}
\begin{align}
\delta^{(1)}G_{00}^{(JJ)}(0,{\bf x};0,0)\Big|_{\delta t\ll |{\bf x}|}
&=\frac{2\pi^{\frac{d+1}{2}}}
{\Gamma\left(\frac{\Delta_{132} }{2}\right)\Gamma\left(\frac{1+d-\Delta_{132} }{2}\right)}
\, {\cal C}\delta \lambda \,\frac{\delta t^{d-\Delta_{132}}}{|{\bf x}|^{2\Delta_{2}}}
\int _{-\infty }^0d\xi\,\frac{f(\xi)}{(-\xi)^{\Delta_{132} +1-d}}\,\notag\\
&+(\Delta_1\leftrightarrow \Delta_2)\,.
\label{fast_quench_J0_J0_result}\\
\delta^{(1)} G_{00}^{(JJ)}(0,{\bf x};0,0)\Big|_{|{\bf x}|\ll \delta t}&=
\frac{2\pi^\frac{d+1}{2}}
{\Gamma\left(\Delta_{3}\right)\Gamma\left(\frac{1+d-2\Delta_{3}}{2}\right)}
\, c_1\,\delta\lambda\,\frac{\delta t^{d-2\Delta_3}}{|{\bf x}|^{\Delta_{123}}} \int _{-\infty}^0d\xi \frac{ f(\xi)}{(-\xi)^{2\Delta_{3}-d+1}}\,.\notag
\end{align}
where
\begin{equation}
\label{C_definition}
{\cal C}=c_1+c_2\,\frac{1-d+\Delta_{132}}{\Delta_{132}}\,.
\end{equation}

The terms in \reef{fast_quench_J0_J0_result} dominate the behavior of the full two-point function in the limit $\delta t\to 0,~\delta\lambda$ fixed.  Moreover, the two-point function is singular in this limit provided that either $\Delta_{132}>d$ or $\Delta_{231}>d$.  In particular, our calculation demonstrates that the scaling of the spatial correlation function of two spin-1 currents flows from $\delta t^{d-2\Delta_3}$ for $x\sim \delta t$ to $\delta t^{d-\Delta_{132}}$ or  $\delta t^{d-\Delta_{231}}$ for $|{\bf x}|\gg \delta t$. From this perspective $G_{00}^{(JJ)}$ scales similarly to its scalar counterpart \cite{Dymarsky:2017awt}. The precise transmutation of one scaling into the other is given by the linear response function \reef{J0J0time_separation}. Furthermore, this scalings are manifest in any continuous field theory with UV fixed point if the quench rate $\delta t^{-1}$ is sufficiently rapid, and therefore \reef{fast_quench_J0_J0_result} is universal.  

Note also that if $J^{(1)}_\mu$ is conserved, then according to \reef{cons_const} ${\cal C}$ in \reef{C_definition} vanishes. Thus $\delta^{(1)}G_{00}^{(JJ)}(0,{\bf x};0,0)$ is given by the exchange $(1\leftrightarrow 2)$ term in \reef{fast_quench_J0_J0_result}. Of course, if both currents are conserved, then both terms in \reef{fast_quench_J0_J0_result} vanish. However, in general the case $\Delta_1=\Delta_2$ is not particularly interesting at large separation $\delta t\ll |{\bf x}|$, since it follows from \reef{fast_quench_J0_J0_result} that in this case the linear response function vanishes in the limit $\delta t\to 0$ for relevant deformations ($\Delta_3<d$).

In fact, (\ref{fast_quench_J0_J0_result}) has simple interpretation in terms of OPE. 
Consider first the limit $\delta t\ll |{\bf x}|$. The integrand in (\ref{quenched_current_current_correction_equal_time})
can be written as follows
\bea
\left[
J^{(1)}_{0}(0,{\bf x})J^{(2)}_{0}(0,0),{\cal O}(t',{\bf y})\right]&=&
J^{(1)}_{0}(0,{\bf x})\,\left[
J^{(2)}_{0}(0,0),{\cal O}(t',{\bf y})\right]
\non
&+& \,\left[J^{(1)}_{0}(0,{\bf x})
,{\cal O}(t',{\bf y})\right]J^{(2)}_{0}(0,0)\,.
\label{use}
\eea
By causality we thus conclude that the non-zero contribution to (\ref{quenched_current_current_correction_equal_time}) comes from the regions where ${\cal O}(t',{\bf y})$ is within the light cone of either $J^{(1)}_{0}(0,{\bf x})$ or $J^{(2)}_{0}(0,0)$, otherwise  commutators simply vanish. However, in the limit $|{\bf x}|\gg \delta t$ the domain defined by the overlap of these light cones with the strip where $\lambda(t')\neq 0$ is space-like separated from the third operator insertion (either $J^{(1)}_{0}(0,{\bf x})$ or $J^{(2)}_{0}(0,0)$ in the above expression). Thus to calculate $\delta^{(1)}G_{00}^{(JJ)}(0,{\bf x};0,0)$ in this limit, it is sensible to use the following OPE, see (\ref{current-current-scalar_general})

\begin{equation}
\label{JJ_OPE}
J_\nu^{(2)}(0){\cal O}(x)\sim\frac{1}{N_J}\,
\frac{c_1\,\delta_\nu^\mu\,x^2+c_2\,x_\nu \,x^\mu}{\left(x^2\right)^{\frac{\Delta_{231}}{2}+1}}\,J_\mu^{(1)}(0)+\dots\,,
\end{equation}
where ellipsis encode various operators which do not contribute to the leading order effect we aim to calculate, $x^\mu=(t'+i\epsilon,~{\bf y})$ and $N_J$ is a normalization constant defined by\,\footnote{The $i\epsilon$ prescription is fixed by the ordering of operators on the left hand side of \reef{JJ_OPE}. \label{epsilon}}
\begin{equation}
\label{N_1_defition}
\langle J^{(i)}_\mu(x)J^{(j)}_\nu(0)\rangle =N_J\, {\delta^{ij}\over (x^2)^{\Delta_{i}}} \( \eta_{\mu\nu} - 2{x_\mu x_\nu\over x^2} \)\,.
\end{equation}
For the temporal component, we thus get
\bea
J_0^{(2)}(0,0){\cal O}(t',{\bf y})&\sim& \frac{1}{N_J}\,
\frac{-(c_1+c_2)\,(t^{\prime}+i\epsilon)^2+c_1|{\bf y}|^2}{\left(-(t'+i\epsilon)^{ 2}+|{\bf y}|^2\right)^\frac{\Delta_{231}+2}{2}}\,J_0^{(1)}(0,0) 
\non
&-& {c_2\over N_J}{(t'+i\epsilon)\, y^i \over \left(-(t'+i\epsilon)^{ 2}+|{\bf y}|^2\right)^\frac{\Delta_{231}+2}{2} } J_i^{(1)}(0,0) 
 +\dots\,,
\eea
Or equivalently,
\bea
[J_0^{(2)}(0,0), {\cal O}(t',{\bf y})]&=&\frac{2i}{N_J}\,J_0^{(1)}(0,0)\,{\rm Im}\,
\frac{-(c_1+c_2)\,(t^{\prime}+i\epsilon)^2+c_1|{\bf y}|^2}{(-(t'+i\epsilon)^{ 2}+|{\bf y}|^2)^\frac{\Delta_{231}+2}{2}}
\non
&-& \frac{2i}{N_J}\,J_i^{(1)}(0,0)\,{\rm Im}\,
{c_2\,(t'+i\epsilon)\, y^i \over \left(-(t'+i\epsilon)^{ 2}+|{\bf y}|^2\right)^\frac{\Delta_{231}+2}{2} } 
\dots\,,
\label{J_0_J_0_fast_quench_OPE_commutator}
\eea
Substituting \reef{use} and \reef{J_0_J_0_fast_quench_OPE_commutator} into
\reef{quenched_current_current_correction_equal_time}, and carrying out the integral over ${\bf y}$ results in the first equation in \reef{fast_quench_J0_J0_result}.

In the opposite regime, $|{\bf x}|\ll \delta t$, the appropriate OPE is
\begin{equation}
\label{J_J_OPE}
J_\mu^{(1)}(0)J_\nu^{(2)}(x)\sim {1 \over N_\mathcal{O}\, (x^2)^{\Delta_{123}\over 2} } 
\Big(c_1\,\eta_{\mu\nu}-(2\,c_1+c_2)\frac{x_{\mu}\,x_{\nu}}{x^2}\Big)\,{\cal O}(0) +\dots\,,
\end{equation}
where $x^\mu=(t'+i\epsilon,~{\bf x})$ and $N_\mathcal{O}$ is the normalization constant defined by
\begin{equation}
\langle {\cal O}(x){\cal O}(y)\rangle =\frac{N_\mathcal{O}}{|x-y|^{2\Delta_{3}}}\,.
\end{equation}
For temporal components at equal times \reef{J_J_OPE} simplifies  
\begin{equation}
\label{J_0J_0_OPE}
J_0^{(1)}(0,{\bf x})J_0^{(2)}(0,0)\sim -\frac{c_1}{N_\mathcal{O}\,|{\bf x}|^{\Delta_{123}}}\,{\cal O}(0)+\dots\,.
\end{equation}
Plugging it into \reef{quenched_current_current_correction_equal_time} we see that in the limit $|{\bf x}|\ll \delta t$ the linear response function $\delta^{(1)}G_{00}^{(JJ)}(t,{\bf x};t,0)$ reduces to $\delta^{(1)} \langle {\cal O}(0)\rangle$. Hence, using
 \cite{Dymarsky:2017awt} 
\begin{equation}
\delta^{(1)} \langle {\cal O}(0)\rangle =-\frac{2\pi^\frac{d+1}{2}\,N_\mathcal{O}}{\Gamma(\Delta_{3})
\Gamma\left(\frac{d-2\Delta_{3}+1}{2}\right)}\,\delta\lambda\,\int_{-\infty}^0 dt'\,\frac{f(t'/\delta t)}{(-t')^{2\Delta_{3}-d+1}}\,,
\end{equation}
we recover the second expression in (\ref{fast_quench_J0_J0_result}).

The correlation function between the spatial components of the currents can be calculated in a similar way. This time
\begin{equation}
\label{quenched_spatial_polarization_current_correction}
\delta^{(1)}G_{ij}^{(JJ)}(t,{\bf x};t,0)=i\,\int _{-\infty}^tdt'\,\lambda(t')\int d^{d-1}{\bf y}\,\left\langle \left[
{\cal O}(t',{\bf y}),J^{(1)}_{i}(t,{\bf x})J^{(2)}_{j}(t,0)\right]\right\rangle_\mt{CFT}\,,
\end{equation}
can be written in terms of the integrals (\ref{Japx}), (\ref{Japx2.1}), (\ref{Japx2}) (see Appendix \ref{master}) as follows
\begin{align}
\label{J_iJ_j_correlator}
\delta^{(1)}G_{ij}^{(JJ)}&(t,{\bf x};t,0)=\frac{2}{|{\bf x}|^{\Delta_{123}}}
\,{\rm Im}\,\int _{-\infty}^t dt'\,\lambda(t')
\bigg( c_2 \, |{\bf x}|^2  J_{ij}(t-t'',{\bf x};\Delta_{132}+2,\Delta_{231}+2,d)\notag\\
&+\left. c_2 \, x_ix_jJ(t-t'',{\bf x};\Delta_{132}+2,\Delta_{231},d)-c_2\, x_j  J_i(t-t'',{\bf x};\Delta_{132}+2,\Delta_{231},d)\right.\notag\\
&+\left. c_2\, x_i  J_j(t-t'',{\bf x};\Delta_{132},\Delta_{231}+2,d)- c_2 \, x_i|{\bf x}|^2 J_j(t-t'',{\bf x};\Delta_{132}+2,\Delta_{231}+2,d)\right.\notag\\
&+\left.
\left(c_1\, \delta_{ij}-\frac{(2\, c_1+c_2)x_ix_j}{|{\bf x}|^2}\right)\,
J(t-t',{\bf x};\Delta_{132},\Delta_{231},d)\right)\,.
\end{align}
Setting $t=0$ and substituting (\ref{quench_profile}), we arrive at 
\begin{align}
\delta^{(1)}G_{ij}^{(JJ)}(0,{\bf x};0,0)\Big|_{\delta t\ll |{\bf x}|}&=
\frac{2\pi^\frac{d+1}{2}}{\Gamma\left(\frac{\Delta_{132} }{2}\right)
\Gamma\left(\frac{1+d-\Delta_{132} }{2}\right)}\,\delta\lambda\,
{\cal C}^{(1)}_{ij}\,
\frac{\delta t^{d-\Delta_{132}}}{|{\bf x}|^{2\Delta_2}}\int _{-\infty}^0d\xi\,\frac{f(\xi)}{(-\xi)^{\Delta_{132} +1-d}}\notag\\
&+(\Delta_1\leftrightarrow \Delta_2)\,.
\label{fast_quench_Ji_Jj_result}\\
\delta^{(1)} G_{ij}^{(JJ)}(0,{\bf x};0,0)\Big|_{|{\bf x}|\ll \delta t}&=
\frac{-2\pi^\frac{d+1}{2}}
{\Gamma\left(\Delta_{3}\right)\Gamma\left(\frac{d+1-2\Delta_{3}}{2}\right)}\,
\delta\lambda\,{\cal C}^{(2)}_{ij}\,\frac{\delta t^{d-2\Delta_3}}{|{\bf x}|^{\Delta_{123}}} \int _{-\infty}^0d\xi \frac{f(\xi)}{(-\xi)^{2\Delta_{3}-d+1}}\,,\notag
\end{align}
where 
\begin{align}
{\cal C}^{(1)}_{ij}(\hat{\bf x})=\(\frac{c_2}{\Delta_{231}} + c_1\)\,\left(\frac{2x_ix_j}{|{\bf x}|^2}-\delta_{ij}\right)\,,
\quad {\cal C}^{(2)}_{ij}(\hat{\bf x})=c_1\(\delta_{ij}-2\,\frac{x_ix_j}{|{\bf x}|^2}\)-c_2\,\frac{x_ix_j}{|{\bf x}|^2}\,.
\end{align}

As before, the scaling behavior \reef{fast_quench_Ji_Jj_result} is universal and can be derived using the OPE technique. The full linear response function \reef{J_iJ_j_correlator} interpolates between the scalings in two extreme limits. For instance, the limit $\delta t\ll |{\bf x}|$ is reproduced using \reef{JJ_OPE} with spatial $\nu$. In particular, this time we get ($t'<0$)
\bea
\label{J_i_J_j_fast_quench_OPE_commutator}
[J_j^{(2)}(0,0),{\cal O}(t',{\bf y})]&=&\frac{2i}{N_J}J_k^{(1)}(0,0)\,\theta(t^{\prime 2}-|{\bf y}|^2)
\frac{c_1\,\delta_j^k\,(-t'^{\,2}+|{\bf y}|^2)+c_2\,y_j\,y^k}{(t^{\prime 2}-
|{\bf y}|^2)^\frac{\Delta_{231}+2}{2}}
\sin\left(\pi\frac{\Delta_{231}}{2}\right)
\non
\\
&+& \frac{2i}{N_J}J_0^{(1)}(0,0)\,\theta(t^{\prime 2}-|{\bf y}|^2)
\frac{c_2\,y_j\,t'}{(t^{\prime 2}-
|{\bf y}|^2)^\frac{\Delta_{231}+2}{2}}
\sin\left(\pi\frac{\Delta_{231}}{2}\right) +\ldots .
\nonumber
\eea
Substituting this commutator into
\bea
&&\delta^{(1)}G_{ij}^{(JJ)}(0,{\bf x};0,0)=-i\,\int _{-\infty}^0dt'\,\lambda(t')\int d^{d-1}{\bf y}\,\left\langle
J^{(1)}_{i}(0,{\bf x})\,\left[
J^{(2)}_{j}(0,0),{\cal O}(t',{\bf y})\right]\right\rangle
\non
&-&i\,\int _{-\infty}^0dt'\,\lambda(t')\int d^{d-1}{\bf y}\,\left\langle
\left[
J^{(1)}_{i}(0,{\bf x}),{\cal O}(t',{\bf y})\right] J^{(2)}_{j}(0,0) \right\rangle ~,
\label{J_i_J_j_fast_quench_OPE_quench_correction}
\eea
and integrating over ${\bf y}$, we recover the first equation in (\ref{fast_quench_Ji_Jj_result}).

To compute the opposite limit, $|{\bf x}|\ll \delta t$, we repeat essentially the same steps as in the calculation of $\delta^{(1)} G_{00}^{(JJ)}(0,{\bf x},0,0)$. The OPE \reef{J_J_OPE} takes the form
\begin{equation}
J_i^{(1)}(0,{\bf x})J_j^{(2)}(0,0)= \frac{ {\cal C}^{(2)}_{ij}(\hat{\bf x}) }{N_\mathcal{O}|{\bf x}|^{\Delta_{123}}}
\,{\cal O}(0)+ \ldots\,.
\end{equation}
Now one can repeat the steps following  \reef{J_0J_0_OPE} to verify the second equation in (\ref{fast_quench_Ji_Jj_result}).

Finally, let us study the two-point function (\ref{quenched_current_current_correction}) with currents inserted at different instants in time, but at the same point in space. We set ${\bf x}=0$ since the quench protocol respects tranlational symmetry. Furthermore, for simplicity we focus on the temporal components only and define
\begin{equation}
\label{Ti_definition}
T_i=t_i-t'-i\epsilon\,,\quad i=1,2\,.
\end{equation}
The essential ingredient in the calculation is encoded in the following integral
\begin{equation}
\label{U_definition}
U(T_1,T_2;\delta _1,\delta _2,d)=
\int d^{d-1}{\bf y}\,\frac{1}{(-T_1^2+|{\bf y}|^2)^{\frac{\delta_1}{2}}
(-T_2^2+|{\bf y}|^2)^{\frac{\delta_2}{2}}}\,,
\end{equation}
which can be evaluated in terms of the hypergeometric functions
\begin{align}
U&=(-T_2^2)^\frac{d-1-\delta_1-\delta_2}{2}\pi^\frac{d-1}{2}\left(\frac{
\Gamma\left(\frac{\delta_2-d+1}{2}\right)}
{\Gamma\left(\frac{\delta_2}{2}\right)}\left(\frac{-T_2^2}
{-T_1^2}\right)^\frac{\delta_1}{2}{}_2F_1\left(\frac{d-1}{2},\frac{\delta_1}{2},
\frac{d+1-\delta_2}{2};\frac{T_2^2}{T_1^2}\right)\right.\\
&+\left.\frac{\Gamma\left(\frac{d-1-\delta_2}{2}\right)
\Gamma\left(\frac{1-d+\delta_1+\delta_2}{2}\right)}
{\Gamma\left(\frac{d-1}{2}\right)\Gamma\left(\frac{\delta_1}{2}\right)}\left(\frac{-T_2^2}
{-T_1^2}\right)^\frac{1+\delta_1+\delta_2-d}{2}{}_2F_1\left(\frac{\delta_2}{2},\frac{1-d+\delta_1+\delta_2}{2},
\frac{3-d+\delta_2}{2};\frac{T_2^2}{T_1^2}\right)\right)\,.\notag
\end{align}
Moreover, it simplifies if we set $t_2=0$ and take the limit of fast quenches, $t_1\gg \delta t$, while assuming $\delta_2>d$ 
\begin{align}
\label{U_large_t_leading}
U(T_1,T_2;\delta _1,\delta _2,d)\Big|_{t_1\gg\delta t}=(-T_2^2)^\frac{d-1-\delta_1-\delta_2}{2}\pi^\frac{d-1}{2}\frac{
\Gamma\left(\frac{\delta_2-d+1}{2}\right)}
{\Gamma\left(\frac{\delta_2}{2}\right)}\left(\frac{-T_2^2}
{-T_1^2}\right)^\frac{\delta_1}{2}\,.
\end{align}
In this case the linear response function takes the form
\begin{align}
\delta ^{(1)}G_{\mu\nu}^{(JJ)}(t_1,0,0,0)\Big|_{t_1\gg\delta t}&=2\pi^\frac{d-1}{2}\delta\lambda
\left(c_1+c_2\,\frac{1-d+\Delta_{231}}{\Delta_{231}}\right)
\frac{\Gamma\left(\frac{\Delta_{231}-d+1}{2}\right)}
{\Gamma\left(\frac{\Delta_{231}}{2}\right)}\,e^{-i\pi\,\frac{\Delta_{123}}{2}}\\
&\times\, c_1\,\frac{\delta t^{d-\Delta_{231}}}{t_1^{2\Delta_{1}}}
\sin\left(\pi\,\frac{d-1-2\Delta_{3}}{2}\right)
\int _{-\infty}^0d\xi \,\frac{f(\xi)}{(-\xi)^{\Delta_{231}-d+1}}\notag\\
&-2\pi^\frac{d-1}{2}\,\delta\lambda\,\left(c_1+c_2\,\frac{1-d+\Delta_{231}}{\Delta_{231}}\right)\frac{\Gamma\left(\frac{\Delta_{231}-d+1}{2}\right)}
{\Gamma\left(\frac{\Delta_{231}}{2}\right)}e^{i\pi\frac{d-1-2\Delta_{2}}{2}}\notag\\
&\times\, c_1\,
\frac{\delta t^{d-\Delta_{231}}}{t_1^{2\Delta_1}}
\sin\left(\frac{\pi \Delta_{132}}{2}\right)\,
\int _0^\infty d\xi\frac{ f(\xi)}{\xi ^{\Delta_{231}-d+1}}\,.\notag
\end{align}

\section{Quenched fermions}
\label{sec:fermions}

In this section we study quantum quenches in the presence of Dirac field $\psi$. Our conventions are reviewed in Appendix \ref{ffO}. We start from considering the linear response of the equal time two-point correlation function 
\begin{equation}
G^{(\psi\psi)}(t,{\bf x};t, 0)\equiv\left\langle \psi_1(t,{\bf x})\bar\psi_2(t,0)\right\rangle\,.
\end{equation}
To ensure validity and universality of the calculations, we focus on the regime when the separation, $|{\bf x}|$, time of observation, $t$, and the duration of quench, $\delta t$, are much smaller than any physical scale inherent to the system or its state. The scaling dimensions of the Dirac fields are denoted by $\Delta_1$ and $\Delta_2$ respectively, whereas the scaling dimension of the deformation is denoted by $\Delta_3$. The linear response of the above correlation function is given by
\begin{equation}
\label{quenched_fermion_fermion_correction}
\delta^{(1)}G^{(\psi\psi)}(t,{\bf x};t, 0)=i\,\int _{-\infty}^tdt'\,\lambda(t')\int d^{d-1}{\bf y}\,\left\langle \left[
{\cal O}(t',{\bf y}),\psi_1(t,{\bf x})\bar\psi_2(t,0)\right]\right\rangle_\mt{CFT}\,.
\end{equation}

As usual, the ordering of operators within the three-point function on the right hand side is achieved by adding a small imaginary component to the Lorentzian time of the operators.\footnote{An operator that is to the `left' of another should have smaller imaginary part.} Using the three-point function
(\ref{fermion-fermion-scalar_general}) derived in Appendix \ref{ffO}, we get
\bea
  &&\langle \left[ {\cal O}(t',{\bf y}),\psi_1(t,{\bf x})\bar\psi_2(t,0)\right] \rangle 
  =  
  \non
  &&\quad\quad \, \frac{ b_1 \, x_j \, \gamma^j }
  { |{\bf x}|^{\Delta_{123}+1} \big({\bf y}^2-(t'-t-i\epsilon)^2\big)^{\Delta_{231}\over 2} \big(({\bf y}-{\bf x})^2-(t'-t-i\epsilon)^2 \big)^{\Delta_{132}\over 2} } 
  \non\non\non
  &&\quad\quad +\, \frac{ b_2 \big( (t'-t-i\epsilon)\gamma^0 + (x-y)_j\gamma^j\big)\big((t'-t-i\epsilon)\gamma^0- y_j\gamma^j  \big)} 
  { |{\bf x}|^{\Delta_{123}} \big({\bf y}^2-(t'-t-i\epsilon)^2\big)^{\Delta_{231}+1\over 2} \big(({\bf y}-{\bf x})^2-(t'-t-i\epsilon)^2 \big)^{\Delta_{132}+1\over 2} } 
 \non\non
 &&\quad\quad - ~ (\epsilon\to -\epsilon) ~.
\eea
where $b_{1,2}$ are constants. In particular, we find it convenient to split the linear response term into two parts proportional to $b_1$ and $b_2$ respectively and evaluate them separately,
\begin{equation}
\label{two_contributions_to_fermion}
\delta^{(1)}G^{(\psi\psi)}(t,{\bf x},t,0)\rangle=b_1\,\delta^{(1)}_1G^{(\psi\psi)}(t,{\bf x},t,0)
+b_2\,\delta^{(1)}_2G^{(\psi\psi)}(t,{\bf x},t,0)\,.
\end{equation}
Thus, for instance, one can write 
\begin{equation}
\label{delta_1_psi_psi_correlator}
\delta^{(1)}_1G^{(\psi\psi)}(t,{\bf x};t,0)=
-2\, \frac{\gamma^ix_i}{|{\bf x}|^{\Delta_{123}+1}}\,  {\rm Im}\, \int _{-\infty}^t dt'\,\lambda(t') J(t'-t,{\bf x};\Delta_{132},
\Delta_{231},d)\,,
\end{equation}
where $J(t'-t,{\bf x};\Delta_{132},\Delta_{231},d)$ is defined in \reef{Japx} and evaluated in Appendix \ref{master}, see \reef{J}.
Similarly, using (\ref{hat_J_i_integral}), (\ref{hat_J_ij_integral}) yields
\begin{align}
\label{delta_2_psi_psi_correlator}
&\delta^{(1)}_2G^{(\psi\psi)}(t,{\bf x};t,0)\\
&= 2\,\int _{-\infty}^t dt'\,\lambda(t')\,\frac{(t-t')x_i\gamma^i\gamma^0+(t-t')^2}
{|{\bf x}|^{\Delta_{123} } } \,{\rm Im}\,
 J(t'-t,{\bf x};\Delta_{132}+1,
\Delta_{231}+1,d)\notag\\
&- \,\frac{2\,\delta^{ij}}{|{\bf x}|^{\Delta_{123}}} \,{\rm Im}   \int _{-\infty}^tdt'\,\lambda(t')
J_{ij}(t'-t,{\bf x};\Delta_{132}+1,
\Delta_{231}+1,d)\notag\\
&+2 \, \frac{x_j\gamma^j\gamma^i}{|{\bf x}|^{\Delta_{123}}}
\,{\rm Im} \int _{-\infty}^tdt'\,\lambda(t')\, J_i(t'-t,{\bf x};\Delta_{132}+1,
\Delta_{231}+1,d)\,,\notag
\end{align}
where $J_i$ and $J_{ij}$ are evaluated in Appendix \ref{master}. Both can be written in terms of $J$.

For the quench protocol \reef{quench_profile} and $\delta t\ll t\ll \ell$, the above linear response functions are proportional to $\delta t$ and therefore vanish in the limit $\delta t \to 0$. Thus, at late times dynamics of the system is governed by the non-linear corrections. However, the scaling structure is rich and universal at early times $t\sim \delta t$. Setting for simplicity $t=0$, and substituting the quench profile (\ref{quench_profile}), we find 
(where $\hat x_i=x_i/|{\bf x}|$)
\bea
&&\delta^{(1)}_1G^{(\psi\psi)}(0,{\bf x};0,0)\Big|_{|{\bf x}|\gg \delta t}\simeq{-2\pi^\frac{d+1}{2} \over \Gamma\left(\frac{\Delta_{132} }{2}\right) \Gamma\left(\frac{d-\Delta_{132} +1 }{2}\right) }\,
{\delta\lambda\over \delta t^{\Delta_{132}-d} }\,
\frac{\gamma^i \hat x_i}{|{\bf x}|^{2\Delta_{2} }} 
 \int _{-\infty}^0d\xi\,{f(\xi)\over (-\xi)^{\Delta_{132}-d+1} }
\non
\non
&&\quad\quad\quad\quad\quad\quad\quad\quad\quad\quad\quad\quad+(1\leftrightarrow 2)\,,
\label{fermion_correlator_fast_quench_first}
\\
\non
&&\delta^{(1)}_1 G^{(\psi\psi)}(0,{\bf x},0,0)\Big|_{|{\bf x}|\ll \delta t}\simeq
\frac{-2 \pi^\frac{d+1}{2}}
{\Gamma\left(\Delta_{3}\right) \Gamma\left(\frac{d-2\Delta_3+1}{2}\right)} \, {\delta\lambda\over \delta t^{\, 2\Delta_3-d}}
\frac{\gamma^i \hat x_i}{|{\bf x}|^{\Delta_{123}} } \int _{-\infty}^0d\xi 
\frac{f(\xi)}{(-\xi)^{2\Delta_{3}-d+1}}\,.
\nonumber
\eea
Similarly, 
\bea
&&\delta^{(1)}_2G^{(\psi\psi)}(0,{\bf x},0,0)\Big|_{|{\bf x}|\gg \delta t}\simeq{2\pi^\frac{d+1}{2}\over \Gamma\left(\frac{\Delta_{132}+1 }{2}\right) \Gamma\left(\frac{d- \Delta_{132}}{2}\right)}
\,{\delta\lambda\over \delta t^{\Delta_{132}-d}}\,
\frac{\gamma^i\gamma^0 \hat x_i}{|{\bf x}|^{2\Delta_2}}
 \int _{-\infty}^0d\xi\,{f(\xi)\over (-\xi)^{\Delta_{132}-d+1} }
\non
\non
&&\quad\quad\quad\quad\quad\quad\quad\quad\quad\quad\quad\quad+(1\leftrightarrow 2)\,,
\label{fermion_correlator_fast_quench_second}
\\
\non
&&\delta^{(1)}_2 G^{(\psi\psi)}(0,{\bf x},0,0)\Big|_{|{\bf x}|\ll \delta t}\simeq {-2\pi^\frac{d+1}{2}\over \Gamma\left(\Delta_3\right) \Gamma\left({d-2\Delta_3+1\over 2} \right)}
\,{\delta\lambda\over \delta t^{2\Delta_3-d}}\,
\frac{1}{|{\bf x}|^{\Delta_{123}} } \int _{-\infty}^0d\xi\frac{ f(\xi)}{(-\xi)^{2\Delta_{3}-d+1}}\,.
\nonumber
\eea

One can understand (\ref{fermion_correlator_fast_quench_first}), (\ref{fermion_correlator_fast_quench_second})
using the OPE approach. For instance, to recover the results in the regime $|{\bf x}|\gg \delta t$, we first observe that $\delta^{(1)}G^{(\psi\psi)}(0,{\bf x}; 0,0)$ can be written as follows 
\bea
\delta^{(1)}G^{(\psi\psi)}(0,{\bf x};0,0)&=&-i\,\int _{-\infty}^0dt'\,\lambda(t')\int d^{d-1}{\bf y}\,\left\langle
\psi_1(0,{\bf x})\,\left[
\bar\psi_2(0,0),{\cal O}(t',{\bf y})\right]\right\rangle
\non
&-&i\,\int _{-\infty}^0dt'\,\lambda(t')\int d^{d-1}{\bf y}\,\left\langle
\,\left[\psi_1(0,{\bf x}),{\cal O}(t',{\bf y})\right] \bar\psi_2(0,0)\right\rangle\,.
\eea
Obviously, causality compels ${\cal O}(t',{\bf y})$ to run within the light cone of either $\psi_1(0,{\bf x})$ or $\bar\psi_2(0,0)$ to ensure the commutators do not vanish. However, in the limit $|{\bf x}|\gg \delta t$ the domain defined by the overlap of these light cones with the region where $\lambda(t')\neq 0$ is space-like separated from the third operator insertion (either $\psi_1(0,{\bf x})$ or $\bar\psi_2(0,0)$ in the above correlation function). Thus to calculate $\delta^{(1)}G^{(\psi\psi)}(0,{\bf x};0,0)$ in this limit, it is sensible to use the following OPE\footnote{The $i\epsilon$ is introduced to match the ordering of operators on the left hand side, whereas the Dirac fields are normalized as follows
\be
 \langle \psi_i(x) \bar\psi_j(0) \rangle = \delta_{ij}\,N_\psi\, {\slashed x \over (x^2)^{\Delta_i+1/2}}~.
\ee
}, see (\ref{fermion-fermion-scalar_general})
\begin{equation}
\label{psi_O_ope_large_x}
\bar\psi_2(0,0){\cal O}(t',{\bf y})\sim{\bar\psi_1(0,0)\over N_\psi}\,
\Big(b_1\,\frac{1}{\(-(t'+i\epsilon)^2+|{\bf y}|^2\)^\frac{\Delta_{231}}{2}}+
b_2\,\frac{-(t'+i\epsilon)\gamma^0+y_i\,\gamma^i}{(-(t'+i\epsilon)^2+|{\bf y}|^2)^\frac{\Delta_{231}+1}{2}}\Big)+\ldots\, ,
\end{equation}
and similarly for $\psi_1(0,0){\cal O}(t',{\bf y})$. In particular, the commutator for $t'<0$ takes the form  
\begin{align}
\label{psi_psi_fast_quench_OPE_commutator}
[\bar\psi_2(0,0),{\cal O}(t',{\bf y})]&\sim -{2\,i\, b_1\over N_\psi} \, \bar\psi_1(0,0)\,\frac{\theta(t^{\prime 2}-|{\bf y}|^2)}
{(t^{\prime 2}-|{\bf y}|^2)^\frac{\Delta_{231}}{2}}\,\sin\left(\pi\,\frac{\Delta_{231}}{2}\right)\\
&+{2\,i\,b_2\over N_\psi} \, \bar\psi_1(0,0)\,(t'\gamma^0-y_i\gamma^i)\,\frac{\theta(t^{\prime 2}-|{\bf y}|^2)}
{(t^{\prime 2}-|{\bf y}|^2)^\frac{\Delta_{231}+1}{2}}\,\sin\left(\pi\,\frac{\Delta_{231}+1
}{2}\right) + \ldots\,.\notag
\end{align}
Plugging it into the expression for $\delta^{(1)}G^{(\psi\psi)}(0,{\bf x};0,0)$, carrying out the integrals over ${\bf y}$ and simplifying the resulting expression gives (\ref{fermion_correlator_fast_quench_first}), (\ref{fermion_correlator_fast_quench_second}).

In the opposite limit, $|{\bf x}|\ll \delta t$, one should use a different OPE to calculate $\delta^{(1)}G^{(\psi\psi)}(0,{\bf x};0,0)$, namely
\begin{align}
\psi_1(0,{\bf x})\bar\psi_2(0,0)\sim\left(b_1\,\frac{\gamma_ix^i}{|{\bf x}|}+b_2\right)\,\frac{1}{N_\mathcal{O}}\,
\frac{1}{|{\bf x}|^{\Delta_{123}}}\,{\cal O}(0)+\ldots \,,
\end{align}
which also follows from (\ref{fermion-fermion-scalar_general}). It is now a straightforward calculation to show that the final answer is consistent with (\ref{fermion_correlator_fast_quench_first}), (\ref{fermion_correlator_fast_quench_second}).

Consider now the linear response of the fermionic two-point function with operator insertions at different instants in time, but at the same point in space. Analogously to (\ref{quenched_current_current_correction})
we have this time
\begin{align}
\label{quenched_fermion_fermion_different_times}
\delta^{(1)}G^{(\psi\psi)}&(t_{1},0;t_{2},0)=i\int _{-\infty}^{t_2}dt'\,\lambda(t')\int d^{d-1}{\bf y}
\langle[{\cal O}(t',{\bf y}),\psi_1(t_1,0)\psi_2 (t_2,0)]\rangle_\mt{CFT}\notag\\
&+i\int _{t_2}^{t_1}dt'\,\lambda(t')\int d^{d-1}{\bf y}
\langle[{\cal O}(t',{\bf y}), \psi_1(t_1,0)]\psi_2 (t_2,0)\rangle_\mt{CFT}\,.
\end{align}
Repeating exactly the same steps as in the previous section, \eg setting for simplicity $t_2=0$ and considering the limit $t_1\gg\delta t$, we get the following result
\begin{align}
&\delta ^{(1)}G^{(\psi\psi)}(t_1,0,0,0)\Big|_{t_1\gg\delta t}=-2\pi^\frac{d-1}{2}\delta\lambda\,
\,e^{-i\pi\,\frac{\Delta_{123}}{2}}\,\frac{\delta t^{d-\Delta_{231}}}{t_1^{2\Delta_{1}}}
\frac{\sin\left(\pi\,\frac{d-1-2\Delta_{3}}{2}\right)}{\Gamma\left(\frac{\Delta_{231}}{2}\right)}\\
&\times \left(b_1\,\Gamma\left(\frac{\Delta_{231}-d+1}{2}\right)\,\gamma^0
+\frac{2ib_2}{\Delta_{231}}\,\Gamma\left(\frac{\Delta_{231}-d+2}{2}\right)\right)
\int _{-\infty}^0d\xi \,\frac{f(\xi)}{(-\xi)^{\Delta_{231}-d+1}}\notag\\
&-2\pi^\frac{d-1}{2}\delta\lambda\,
\,e^{i\pi\,\frac{d-1-2\Delta_2}{2}}\,\frac{\delta t^{d-\Delta_{231}}}{t_1^{2\Delta_{1}}}
\frac{\sin\left(\pi\,\frac{\Delta_{123}}{2}\right)}{\Gamma\left(\frac{\Delta_{231}}{2}\right)}\notag\\
&\times \left(b_1\,\Gamma\left(\frac{\Delta_{231}-d+1}{2}\right)\,\gamma^0
+\frac{2ib_2}{\Delta_{231}}\,\Gamma\left(\frac{\Delta_{231}-d+2}{2}\right)\right)
\int _{-\infty}^0d\xi \,\frac{f(\xi)}{(-\xi)^{\Delta_{231}-d+1}}\,.\notag
\end{align}

%\section{Discussion}
%\label{sec:discussion}

\acknowledgments  
We thank Anatoly Dymarsky for helpful discussions and comments.  This work is supported by the BSF grant 2016186 and by the "Quantum Universe" I-CORE program of the Israel Planning and Budgeting Committee (grant 1937/12).

\appendix
\section{Projective null cone}

In this Appendix we give a brief outline of the embedding space formalism needed for our calculations. Our presentation makes use and relies on the work by others \cite{Weinberg:2010fx,Costa:2011mg,Rychkov:2016iqz}. 

It is well known that the connected part of the conformal group in a $d$-dimensional Minkowski space can be realized as linear transformations $SO(d,2)$ in $\mathbb{R}^{d,2}$. In particular, if we denote the coordinates of the $d+2$-dimensional embedding space by $X^M$,
($M=+,-,\mu$), then the $d$-dimensional CFT is accommodated on a section of the light cone,
\begin{equation}
 \eta_{MN}X^M X^N = 0\,,\qquad  X^MdX_M=0\,,\label{light_cone}
\end{equation}
parametrized by
\begin{equation}
X^\mu=x^\mu\,,\qquad X^+=f(x^\mu)\,,\qquad X^+X^-=x^2\,,
\end{equation}
where $x^\mu$ are coordinates of the CFT, $X^+=f(x)$
defines the light cone section, and $X^-$ is fixed by the
light cone constraint. We denoted the light-cone coordinates as
\begin{equation}
X^\pm =X^6\pm X^5\,.
\end{equation}

The metric of the ambient space,
\begin{equation}
ds^2=\eta_{MN}dX^MdX^N=-dX^+dX^-+\eta_{\mu\nu}dX^\mu dX^\nu\,,
\label{lightcone}
\end{equation}
determines the induced metric on the light cone section where CFT lives. For a flat section a convenient choice is $f(x)\equiv 1$, in which case the light cone constraint yields $X^-=x^2$. As a result, the $d$-dimensional CFT lives on the subspace of $\mathbb{R}^{d,2}$ defined by
\begin{equation}
X^M(x)=(1,x^2,x^\mu)\,,
\label{embed}
\end{equation}
whereas the conformal group consists of $SO(d,2)$ transformations
\begin{equation}
X^M\rightarrow \Lambda^M_{\;\;N}X^N\,.
\end{equation}

To ensure $X^+=1$ holds after the above linear transformation takes place we supplement it with rescaling $X^M\rightarrow \lambda(x)\,X^M$ of the form
\begin{equation}
\lambda(x)=\left(\Lambda^+_{\;\;+}+\Lambda^+_{\;\;-}x^2+\Lambda^+_{\;\;\mu}x^\mu\right)^{-1}\,.
\end{equation}
Since the light cone constraint $X^MX_M=0$ is invariant under both transformations, such a combination of boost plus scaling defines a diffeomorphism of the subspace (\ref{embed}). In particular, the induced metric remains invariant up to a scale factor. This can be seen from the following sketchy argument
\begin{equation}
dX^2\rightarrow dX^{\prime 2}=d(\lambda(X)X)^2=\lambda(x)^2dX^2\,,
\end{equation}
where the light cone condition (\ref{light_cone}) was used in the last equality.

The primary fields of the CFT correspond to tensors of $SO(d,2)$ living on a light cone and satisfying certain conditions. For instance, a scalar primary $\mathcal{O}(x)$ with scaling dimension $\Delta_\mathcal{O}$ is uplifted to a scalar $\mathcal{O}(X)$ defined on the light cone \reef{lightcone} and satisfying the homogeneity condition $\mathcal{O}(\lambda X)=\lambda^{-\Delta_\mathcal{O}}\,\mathcal{O}(X)$. Similarly, a primary vector field, $J_\mu(x)$, is uplifted to a vector, $J_M(X)$, of $SO(d,2)$  satisfying the homogeneity and transversality conditions
\bea
J_M(\lambda X)=\lambda^{-\Delta_J}\,J_M(X)\,, \quad X^M\,J_M(X)=0\,,
\eea
where $\Delta_J$ stands for the scaling dimension of $J_\mu(x)$. The connection between the fields is provided by\footnote{Note that $X_M$ is projected to zero because of (\ref{light_cone}) and (\ref{tensor_field_projection_definition}). Hence, (\ref{tensor_field_projection_definition}) projects any $J_M(X)$ and $J_M(X)+\alpha X_M$ onto the same vector $J_\mu(x)$. Furthermore, since transversality condition eliminates one of the component of $J_M(X)$, the match between $J_\mu(x)$ and $J_M(X)$ is one-to-one up to $J_M\sim X_M$.}
\begin{equation}
\label{tensor_field_projection_definition}
\mathcal{O}(x)=\mathcal{O}(X)\Big|_{X^M(x)}\, , \quad J_\mu(x)=\frac{\partial X^M}{\partial x^\mu}\, J_M(X)\Big|_{X^M(x)}\,.
\end{equation}

\section{Current-current-scalar correlation function}
\label{appendix:jjO}

In this Appendix we derive the three point function of primary scalar and two spin-1 currents used in the text. This is a particular case of the three point function calculated in \cite{Costa:2011mg}. The corresponding correlator in $\mathbb{R}^{d,2}$ is given by
\begin{equation}
G^{(JJ{\cal O})}_{MN}=\left\langle J_M^{(1)}(X_1)J_N^{(2)}(X_2){\cal O}(X_3)\right\rangle\,.
\end{equation}
Let us denote by $\Delta_{J_1}, \Delta_{J_2}$ and $\Delta_{\cal O}$ the scaling dimensions of $J_M(X_1), J_N(X_2)$ and ${\cal O}(X_3)$ respectively. Based on the homogeneity of operators under rescaling of their argument and simple transformation rule under $SO(d,2)$ group in the ambient space, we deduce that the most general ansatz for $G^{(JJ{\cal O})}_{MN}$ can be written as a product of the scalar three-point function 
\begin{align}
S^{(JJ{\cal O})}(X_1,X_2,X_3)&=\frac{1}{\left(X_1\cdot X_2\right)^\frac{\Delta_{J_1J_2\, {\cal O}}}{2}
\left(X_1\cdot X_3\right)^\frac{\Delta_{J_1\,{\cal O}\, J_2}}{2}
\left(X_2\cdot X_3\right)^\frac{\Delta_{J_2\,{\cal O}\, J_1}}{2}}\,,
\end{align}
where for brevity 
\begin{equation}
\Delta_{ABC}=\Delta_A+\Delta_B-\Delta_C\,,\quad A,B,C=J_1,J_2,{\cal O}\,,
\end{equation}
and the scale-invariant tensor 
\begin{align}
T^{(JJ{\cal O})}_{MN}(X_1,X_2,X_3)=\eta_{MN}&+c_1\,\frac{X_{2\,M}X_{1\,N}}{X_1\cdot X_2}
+c_2\,\frac{X_{3\,M}X_{1\,N}}{X_1\cdot X_3}
+c_3\,\frac{X_{2\,M}X_{3\,N}}{X_2\cdot X_3}\\
&+c_4\,\frac{X_{1\,M}X_{2\,N}}{X_1\cdot X_2}
+ c_5\,\frac{X_{1\,M}X_{3\,N}}{X_1\cdot X_3}
+ c_6\,\frac{X_{3\,M}X_{2\,N}}{X_2\cdot X_3}\\
&+c_7\,\frac{X_{1\,M}X_{1\,N}\,X_2\cdot X_3}{X_1\cdot X_2\, X_1\cdot X_3}\\
&+c_8\,\frac{X_{2\,M}X_{2\,N}\,X_1\cdot X_3}{X_1\cdot X_2\, X_2\cdot X_3}\\
&+c_9\,\frac{X_{3\,M}X_{3\,N}\,X_1\cdot X_2}{X_1\cdot X_3\, X_2\cdot X_3}\,,
\end{align}
where $c_i$'s are arbitrary constants that can be related to each other by imposing transversality and light cone constraints 
\begin{align}
X_n\cdot X_n=0\,,\quad n=1,2,3 \,,\qquad X_n^M\,\frac{\partial X_{n\,M}}{\partial x_i^\mu}=0\,,\label{conic_vector_constraint}\\
X_1^M\,J_M(X_1)=0\,,\qquad X_2^N\,J_N(X_2)=0\,.\label{conic_vector_projection}
\end{align}

Considering that projection is carried out through the use of \reef{embed} and \reef{tensor_field_projection_definition}, we can ignore terms proportional to $c_{4,5,6,7,8}$ as their projection eventually vanishes because of (\ref{conic_vector_constraint}). For the rest of $c$'s (\ref{conic_vector_projection}) results in 
\begin{equation}
c_1=-1-{c\over 2}\,,\quad c_2={c\over 2}\,,\quad c_3={c\over 2}\,,\quad c_9=-{c\over 2}\,.
\end{equation}
Combining altogether gives
\begin{align}
&T^{(JJ{\cal O})}_{MN}(X_1,X_2,X_3)=\eta_{MN}-\frac{X_{2\,M}X_{1\,N}}{X_1\cdot X_2}\\
&+{c\over 2}\,\left(\frac{X_{3\,M}X_{1\,N}}{X_1\cdot X_3}
+\frac{X_{2\,M}X_{3\,N}}{X_2\cdot X_3}
-\frac{X_{2\,M}X_{1\,N}}{X_1\cdot X_2}
-\frac{X_1\cdot X_2\,X_{3\,M}X_{3\,N}}{X_1\cdot X_3\,X_2\cdot X_3}\right)+ \ldots \,,\notag
\end{align}
where ellipsis encode terms which are annihilated by projection \reef{tensor_field_projection_definition}. This result is in full agreement with \cite{Costa:2011mg}.

Furthermore, on on a sub-manifold \reef{embed} where the CFT lives, we have
\begin{align}
\frac{\partial X^M}{\partial x^\mu}&=\left(0,\, 2x_\mu,\,\delta^M_\mu\right)\,,\\
\eta_{ML}\,X_2^M\,\frac{\partial X_1^L}{\partial x_1^\mu}&=x_{2\mu}-x_{1\mu}\,,\\
X_1\cdot X_2&=-\frac{1}{2}\,x_{12}^2\,,\qquad x_{12}^2=|x_1-x_2|^2\,.
\end{align}
Hence, after applying projection \reef{tensor_field_projection_definition} the scalar and tensor parts of the conformal correlation function (\ref{vector_vector_scalar_definition}) take the form \reef{scalar_pt} and \reef{tensor_pt}.

\section{Master integrals}
\label{master}

Here we calculate the integrals encountered in the text numerous times
\be
J(t,{\bf x};\delta_1,\delta_2,d)\equiv \int d^{d-1}{\bf y}\,\frac{1}{(-(t-i\epsilon)^2+({\bf y}-{\bf x})^2)^{\frac{\delta_1}{2}}
(-(t-i\epsilon)^2+{\bf y}^2)^{\frac{\delta_2}{2}}}\,.
\label{Japx}
\ee

Introducing Feynman parameter $u$, and then shifting the integration variable,
${\bf y}\rightarrow {\bf y}+u\,{\bf x}$, we obtain
\begin{align}
J(t,{\bf x};\delta_1,\delta_2,d)&=
\frac{\Gamma\left(\frac{\delta_1+\delta_2}{2}\right)}{\Gamma\left(\frac{\delta_1}{2}\right)
\Gamma\left(\frac{\delta_2}{2}\right)}\int _0^1 du\, u^{\frac{\delta_1}{2}-1}
(1-u)^{\frac{\delta_2}{2}-1}\\
&\times\int d^{d-1}{\bf y}\,
\frac{1}{({-}(t{-}i\epsilon)^2{+}|{\bf y}|^2{+}u(1{-}u)|{\bf x}|^2)^\frac{\delta_1{+}\delta_2}{2}}\,.
\end{align}
The integral over ${\bf y}$ is now straightforward,
\begin{align}
J(t,{\bf x};\delta_1,\delta_2,d)&=\frac{\pi^\frac{d-1}{2}\Gamma\left(\frac{\delta_1+\delta_2-d+1}{2}\right)}
{\Gamma\left(\frac{\delta_1}{2}\right)\Gamma\left(\frac{\delta_2}{2}\right)}\\
&\times \int _0^1 du\, u^{\frac{\delta_1}{2}-1}
(1-u)^{\frac{\delta_2}{2}-1}\,({-}(t{-}i\epsilon)^2{+}u(1{-}u)|{\bf x}|^2)^{\frac{d-1-\delta_1-\delta_2}{2}}\,.
\end{align}
Next we introduce a convenient variable
\begin{equation}
z=\frac{|{\bf x}|^2}{(t-i\epsilon)^2}\,,
\end{equation}
which allows us to rewrite
\begin{align}
J(t,{\bf x},;\delta_1,\delta_2,d)&=\frac{\pi^\frac{d-1}{2}\Gamma\left(\frac{\delta_1+\delta_2-d+1}{2}\right)}
{\Gamma\left(\frac{\delta_1}{2}\right)\Gamma\left(\frac{\delta_2}{2}\right)}
({-}(t{-}i\epsilon)^2)^{\frac{d-1-\delta_1-\delta_2}{2}}\\
&\times \int _0^1 du\, u^{\frac{\delta_1}{2}-1}
(1-u)^{\frac{\delta_2}{2}-1}\,(1{-}u(1{-}u)z)^{\frac{d-1-\delta_1-\delta_2}{2}}\,.
\end{align}
Using definition of the Pochhammer symbol
\begin{equation}
(a)_k=a(a+1)\cdots (a+k-1)\,,\quad (a)_0=1\,,
\end{equation}
its property
\begin{equation}
(a)_{2k}=2^{2k}\left(\frac{a}{2}\right)_k\left(\frac{a+1}{2}\right)_k\,,
\end{equation}
and representation of the generalized hypergeometric function
\begin{equation}
{}_3F_2(a_1,a_2,a_3;b_1,b_2;z)=\sum_{k=0}^\infty
\frac{(a_1)_k(a_2)_k(a_3)_k\,z^k}{(b_1)_k(b_2)_k\, k!}\,,
\end{equation}
we can calculate the integral over $u$
\begin{align}
&\int _0^1 du\, u^{\frac{\delta_1}{2}-1}
(1-u)^{\frac{\delta_2}{2}-1}\,(1{-}u(1{-}u)z)^{\frac{d-1-\delta_1-\delta_2}{2}}\\
&=\sum_{k=0}^\infty\frac{z^k(\frac{\delta_1{+}\delta_2{-}d{+}1}{2})_k}{k!}\,\int _0^1 du\, u^{\frac{\delta_1}{2}-1+k}
(1-u)^{\frac{\delta_2}{2}-1+k}\\
&=\frac{\Gamma\left(\frac{\delta_1}{2}\right)\Gamma\left(\frac{\delta_2}{2}\right)}
{\Gamma\left(\frac{\delta_1+\delta_2}{2}\right)}\,{}_3F_2\left(\frac{\delta_1}{2},\frac{\delta_2}{2},\frac{\delta_1+\delta_2-d+1}{2};
\frac{\delta_1+\delta_2}{4},\frac{\delta_1+\delta_2+2}{4};\frac{z}{4}\right)\,.
\end{align}
Hence,
\begin{align}
J(t,{\bf x};\delta_1,\delta_2,d)&=\frac{\pi^\frac{d-1}{2}\Gamma\left(\frac{\delta_1+\delta_2-d+1}{2}\right)}
{\Gamma\left(\frac{\delta_1+\delta_2}{2}\right)({-}(t{-}i\epsilon)^2)^{\frac{\delta_1{+}\delta_2{-}d{+}1}{2}}}
\label{J}
\\
&\times{}_3F_2\left(\frac{\delta_1}{2},\frac{\delta_2}{2},\frac{\delta_1+\delta_2-d+1}{2};
\frac{\delta_1+\delta_2}{4},\frac{\delta_1+\delta_2+2}{4};\frac{|{\bf x}|^2}{4(t-i\epsilon)^2}\right)\,.\notag
\end{align}
Using this result we can readily evaluate two additional integrals used in the text
\bea
 J_i(t,{\bf x},;\delta_1,\delta_2,d)&\equiv&
\int d^{d-1}{\bf y}\,\frac{y_i}{(-(t-i\epsilon)^2+({\bf y}-{\bf x})^2)^{\frac{\delta_1}{2}}
(-(t-i\epsilon)^2+{\bf y}^2)^{\frac{\delta_2}{2}}}\,,
\label{Japx2.1}\\
 J_{ij}(t,{\bf x},;\delta_1,\delta_2,d)&\equiv&
\int d^{d-1}{\bf y}\,\frac{y_i\,y_j}{(-(t-i\epsilon)^2+({\bf y}-{\bf x})^2 )^{\frac{\delta_1}{2}}
(-(t-i\epsilon)^2+{\bf y}^2)^{\frac{\delta_2}{2}}}\,.
\label{Japx2}
\eea
Indeed, based on the definition \reef{Japx}, we have
\bea
  J_i(t,{\bf x},;\delta_1,\delta_2,d) &=&\(- {1\over \delta_1} {\del\over \del x^i}J(t,{\bf x},;\delta_1,\delta_2,d) \)\Bigg|_{\delta_1\to \delta_2-2\,, ~ \delta_2\to \delta_1} 
  \\
  J_{ij}(t,{\bf x},;\delta_1,\delta_2,d) &=&\({1\over \delta_1(\delta_1+2)} {\del^2 \over \del x^i \del x^j}J(t,{\bf x},;\delta_1,\delta_2,d) \right.
  \non
  && \quad\quad\quad\quad + \left. {\delta_{ij}\over \delta_1+2} J(t,{\bf x},;\delta_1+2,\delta_2,d) \)_{\delta_1\to \delta_2-4\,, ~ \delta_2\to \delta_1} \, .
  \nonumber
\eea
Taking derivatives of \reef{Japx2} and rearranging terms, yields
\begin{align}
 J_i(t,{\bf x},;\delta_1,\delta_2,d)&=x_i\,\frac{\delta_1}{2\pi}\, J(t,{\bf x};\delta_1+2,\delta_2,d+2)\, ,
\label{hat_J_i_integral}
\\
 J_{ij}(t,{\bf x},;\delta_1,\delta_2,d)&=\frac{\delta_{ij}}{2\pi}\, J(t,{\bf x};\delta_1,\delta_2,d+2)
+\frac{\delta_1(\delta_1+2)}{4\pi^2}\,x_i\,x_j\, J(t,{\bf x};\delta_1+4,\delta_2,d+4)\,.
\label{hat_J_ij_integral}
\end{align}

\section{Fermion-fermion-scalar correlation function}
\label{ffO}

In this appendix we use the embedding space formalism to derive the conformal three-point function of two primary Dirac fields $\psi_{1,2}(x)$ and a primary scalar $\mathcal{O}(x)$ in $\mathbb{R}^{d-1,1}$. The scaling dimensions of the fields are denoted by $\Delta_{\psi_1}$,  $\Delta_{\psi_2}$ and $\Delta_\mathcal{O}$ respectively. 

Our analysis closely follows \cite{Weinberg:2010fx}. In particular, we do not impose $X^+=1$ throughout this appendix, and the points of $\mathbb{R}^{d-1,1}$ are identified with the light cone generating rays. The connection between the coordinates of $\mathbb{R}^{d-1,1}$ and $\mathbb{R}^{d,2}$ is provided by the formula
\begin{equation}
x^\mu ={X^\mu\over X^+}~, \quad X^-=X^+\,x^2\,.
\label{coor}
\end{equation}

As in the case of tensor fields with integer spin, the primary spinors $\psi_{1,2}(x)$ are uplifted to Dirac fields $\Psi_{1,2}(X)$ living on the light cone in $\mathbb{R}^{d,2}$ and obeying homogeneity and transversality conditions
\begin{align}
\Psi_{1,2}(\lambda X)&=\lambda^{1/2-\Delta_{\psi_{1,2}}}\,\Psi_{1,2}(X)\,, \quad (X\cdot\Gamma)\Psi_{1,2}(X)=0 ~,
\label{psicon}
\end{align}
where our choice for the representation of gamma matrices, $\Gamma^M$, in $\mathbb{R}^{d,2}$ is \cite{Isono:2017grm}
\begin{equation}
\Gamma^\mu =\left({\gamma^\mu \atop 0}\;{0\atop -\gamma^\mu}\right)\,,\;\;
\mu=0,\dots, d-1\,,\quad \Gamma^{+}=\left({0\atop 0}\;{2\atop 0}\right)\,,\quad
\Gamma^{-}=\left({0\atop -2}\;{0\atop 0}\right)\,,
\label{gamma2}
\end{equation}
with $2^{\[d\over 2\]}\times 2^{\[d\over 2\]}$ matrices $\gamma^\mu, ~ \mu=0,..,d-1$ representing Clifford algebra in $d$-dimensional spacetime.\footnote{In even dimensional space-time there exists the so-called chirality gamma matrix. In $\mathbb{R}^{d-1,1}$ and $\mathbb{R}^{d,2}$, we define them as follows
\bea
\gamma_5&=&i^{2-d\over 2}\prod_{\mu=0}^{d-1} \gamma^\mu~,
\non
 \Gamma_5&\equiv&{(i)^{-{d+2\over 2}}\over 4} \[\Gamma^-,\Gamma^+\]\prod_{\mu=0}^{d-1} \Gamma^\mu=
\left({-1\atop~~\, 0}\;{0\atop 1}\right)\,.
\eea
}

The rows and columns of the supermatrices in \reef{gamma2} will be labelled by $\pm$ index. Thus, for instance, the $2^{{\[d\over 2\]}+1}$-component Dirac field takes the form
\begin{equation}
\Psi=\left({\Psi_+\atop \Psi_-}\right)\,.
\end{equation}
It transforms in a standard way under the generators, $J^{MN}$, of the $SO(d,2)$ group
\begin{equation}
\label{geneal_lorentz_in_embedding_space}
i[J^{MN},\Psi]=(X^N\partial ^M-X^M\partial ^N)\Psi -i\,\mathcal{J}^{MN}\Psi\,,
\end{equation}
where $\mathcal{J}^{MN}=[\Gamma^M,\Gamma^N]/(4i)$ build the Dirac representation of the $SO(d,2)$ Lie algebra.

The spinor $\Psi$ should be related to the Dirac field  $\psi$ in $\mathbb{R}^{d-1,1}$ such that the latter obeys the following transformation rules under the generators of  conformal group
\begin{align}
i[J^{\mu\nu},\psi]&=(x^\nu\partial ^\mu -x^\mu\partial ^\nu)\psi -i\, j^{\mu\nu}\psi\,,
\label{lorentz_transform_projected_spinor}\\
i[P^\mu,\psi]&=-\partial ^\mu\psi\,,
\label{translation_transform_projected_spinor}\\
i[K^\mu,\psi]&=(2x^\mu x^\lambda \partial _\lambda -x^2\partial ^\mu +2\Delta x^\mu)\psi
+2i\, j^{\mu\nu}x_\nu\psi\,,
\label{special_transform_projected_spinor}\\
i[S,\psi]&=(x^\lambda \partial _\lambda +\Delta)\psi\,,\label{scale_transform_projected_spinor}
\end{align}
where $j^{\mu\nu}=[\gamma^\mu,\gamma^\nu]/(4i)$ form the Dirac representation of the  Lorentz group Lie algebra, whereas the generators translations, $P^\mu$,  dilations, $S$, and special conformal transformations, $K^\mu$, are simply related to their counterparts $J^{MN}$
\begin{align}
 P^\mu= J^{+\mu}\,,\quad K^\mu=J^{-\mu}\,,\quad S= \frac{1}{2}\,J^{-+}\,.
 \label{cofgen}
\end{align}

To construct the desired relation between $\Psi$ and $\psi$, we start from defining an auxiliary spinor
\begin{equation}
\zeta=\left({\zeta_+\atop \zeta_-}\right)=(X^+)^{\Delta-\frac{1}{2}}\Psi~.
\end{equation}
According to \reef{psicon} it does not change under scaling, \ie by definition the auxiliary spinor is invariant along the rays that generate the light cone. Hence, $\zeta$ a well-defined function of $x^\mu$, and we can think of it as an object in $\mathbb{R}^{d-1,1}$ satisfying the constraint
\begin{equation}
\label{zeta_projection_constraint}
X_M\,\Gamma^M\,\zeta=0\quad\Rightarrow\quad \left({x_\mu \gamma^\mu \atop 1}\;{-x^2\atop -x_\mu\gamma^\mu}\right)
\left(\zeta_+\atop \zeta _-\right)=0\quad\Rightarrow\quad \zeta_+=x_\mu\gamma^\mu\zeta_- \,.
\end{equation}

The auxiliary spinor $\zeta$ cannot be directly identified with a smaller $\psi$ living in $\mathbb{R}^{d-1,1}$. Furthermore, $\zeta$ does not have the usual commutation relations with the generators of conformal group in $\mathbb{R}^{d-1,1}$. Thus, for instance, using \reef{geneal_lorentz_in_embedding_space} and \reef{cofgen}, gives
\begin{align}
i[P^\mu,\zeta _+]&=-\partial^\mu \zeta _+    +\gamma^\mu \zeta_-\,,\\
i[P^\mu,\zeta _-]&=-\partial^\mu \zeta_-  \,.
\label{translations_zeta_plus_mins}
\end{align}
Similarly,
\begin{align}
i[K^\mu,\zeta _+]&=(2x^\mu x^\lambda\partial _\lambda -x^2\partial^\mu +(2\Delta-1)x^\mu)\zeta_+\,,\\
i[K^\mu,\zeta _-]&=(2x^\mu x^\lambda\partial _\lambda -x^2\partial^\mu +(2\Delta-1)x^\mu)\zeta_- +\gamma^\mu \zeta_+ \,,
\label{special_zeta_plus_mins}
\end{align}
and
\begin{align}
i[S,\zeta _+]&=(x^\lambda \partial_\lambda+\Delta-1)\zeta_+\,,\\
i[S,\zeta _-]&=(x^\lambda \partial_\lambda+\Delta)\zeta_- \,.
\label{scaling_zeta_plus_mins}
\end{align}
However, using the transversality constraint (\ref{zeta_projection_constraint}), we can rewrite (\ref{special_zeta_plus_mins}) as follows
\begin{equation}
i[K^\mu,\zeta _-]=(2x^\mu x^\lambda\partial _\lambda -x^2\partial^\mu +2\Delta x^\mu)\zeta_-+
2i\,j^{\mu\nu}\,x_\nu \zeta_-\,,
\label{special_zeta_mins}
\end{equation}
In particular, it follows from \reef{translations_zeta_plus_mins}, \reef{scaling_zeta_plus_mins} and \reef{special_zeta_mins} that $\zeta_-$ transforms according to \reef{lorentz_transform_projected_spinor}-\reef{scale_transform_projected_spinor} under the conformal group in $\mathbb{R}^{d-1,1}$. Hence, correct identification takes the form 
\begin{equation}
\psi=\zeta_-=(X^+)^{\Delta-\frac{1}{2}}\,\Psi_-\,.
\label{relpsi1}
\end{equation}

Now let us define the Dirac adjoints in $\mathbb{R}^{d-1,1}$ and $\mathbb{R}^{d,2}$ as $\bar\psi\equiv i \psi^\dagger\gamma^0$and $\bar\Psi\equiv\Psi^\dagger \beta$ respectively, where 
\begin{equation}
\beta={i\over 2}\, \Gamma^0\big(\Gamma^+ +\Gamma^-\big)=\left({0\atop i\gamma^0}\;{i\gamma^0\atop 0}\right)\,,
\quad
\beta^{-1}=\beta^\dagger =\beta\,,\quad \beta\Gamma^M \beta=(\Gamma^M)^\dagger\,.
\end{equation}
Thus,
\bea
\bar\Psi&=&\Psi^\dagger \beta=\begin{pmatrix}
      i\Psi_-^\dag\gamma^0  &~ 
      i \Psi_+^\dag\gamma^0 
\end{pmatrix}
\,, \non
\bar\psi&\equiv& i\zeta_-^\dag\gamma^0=(X^+)^{\Delta-\frac{1}{2}}i\Psi_-^\dag\gamma^0=(X^+)^{\Delta-\frac{1}{2}}\bar\Psi_+\, .
\label{relpsi2}
\eea

Next we note that the most general ansatz for the $SO(d,2)$ invariant three-point function in $\mathbb{R}^{d,2}$ is 
\begin{align}
G^{\Psi\Psi\,{\cal O}}(X,Y,Z)&\equiv\langle \Psi_1(X)\bar\Psi_2(Y){\cal O}(Z)\rangle \non
&=C_1+C_2 \, X\cdot \Gamma +C_3\, Y\cdot\Gamma + C_4\,Z\cdot\Gamma\non
&+C_5\,[X\cdot\Gamma, Y\cdot\Gamma]+C_6\, [Y\cdot\Gamma, Z\cdot\Gamma]+C_7\,[X\cdot\Gamma, Z\cdot\Gamma]\\
&+C_8\,(X\cdot\Gamma)\, (Z\cdot\Gamma) \, (Y\cdot\Gamma)\,,
\nonumber
\end{align}
where all $C_i$'s are scalar functions of $X\cdot Y$, $X\cdot Z$, $Y\cdot Z$. The term proportional to $C_8$ is not antisymmetrized to simplify imposing the transversality constraints \reef{psicon} associated with the conical section. Antisymmetrization of this term amounts to simple redefinition of other terms in the ansatz. 

The transversality constraints \reef{psicon} give
\begin{equation}
(X\cdot\Gamma) \, G^{\Psi\Psi\,{\cal O}}(X,Y,Z)=0\,,\qquad   G^{\Psi\Psi\,{\cal O}}(X,Y,Z)\, (Y\cdot\Gamma)=0 ~.
\end{equation}
They lead to a set of relations obeyed by various $C_i$'s. To display these relations explicitly, we use the following identities 
\bea
(X\cdot \Gamma) \, (Y\cdot\Gamma)(X\cdot\Gamma) &=&2(X\cdot Y)\, (X\cdot \Gamma)\,,\non
(X\cdot \Gamma) \, (Y\cdot\Gamma)&=&X\cdot Y+\frac{1}{2}[X\cdot \Gamma , Y\cdot\Gamma]\,,\non
\Gamma ^M\,[\Gamma^N,\Gamma^K]&=&2\eta^{MN}\Gamma^K-2\eta^{MK}\Gamma^N+
\frac{1}{3}\Gamma^{[M}\Gamma^N\Gamma^{K]}\,,\non
\[\Gamma^M,\Gamma^N\]\Gamma^K&=&2\eta^{NK}\Gamma^M-2\eta^{MK}\Gamma^N+
\frac{1}{3}\Gamma^{[M}\Gamma^N\Gamma^{K]}\,,
\nonumber
\eea
where the square brackets around the indices stand for antisymmetrization.\footnote{We do not include $1/3!$ factor in the definition of antisymmetrization.} In particular, we obtain
\bea
0&=&(X\cdot\Gamma) \, G^{\Psi\Psi\,{\cal O}}(X,Y,Z)=C_3\, X\cdot Y+C_4\, X\cdot Z
\non
&+&\big(C_1-2C_5\, X\cdot Y-2C_7\, X\cdot Z\big)\, (X\cdot \Gamma)-2C_6\,(X\cdot Z)\,(Y\cdot \Gamma)+2C_6\, (X\cdot Y)\, (Z\cdot \Gamma)
\non
&+&\frac{1}{2}\Big( C_3\,X_M Y_N+ C_4\,X_M Z_N\Big) \Gamma^{[M}\Gamma^{N]}
+\frac{1}{3}C_6 X_MY_NZ_K\Gamma^{[M}\Gamma^{N}\Gamma^{K]}~.
\eea
Recalling now that $\mathbb{I}, ~ \Gamma^M$ and antisymmetrized products of gamma matrices are linearly independent, yields 
\begin{equation}
C_3=C_4=C_6=0\,,\quad  C_1=2C_5\, X\cdot Y+2C_7\, X\cdot Z\,.
\end{equation}
Similarly,
\bea
0=G^{\Psi\Psi\,{\cal O}}(X,Y,Z)\,Y\cdot\Gamma &=&C_2\,X\cdot Y+2C_7\,(Y\cdot Z)\, (X\cdot\Gamma)
\non
&&+\big(C_1-2C_5\,X\cdot Y\big)\,
Y\cdot\Gamma-2C_7(X\cdot Y)(Z\cdot\Gamma)
\non
&&+\frac{1}{2}C_2\,X_MY_N\, \Gamma^{[M}\Gamma^{N]}+\frac{1}{3}C_7X_MZ_NY_K\Gamma^{[M}\Gamma^N\Gamma^{K]} ~.
\nonumber
\eea
Or equivalently,
\begin{equation}
C_2=C_7=0\,,\quad  C_1=2C_5\, X\cdot Y\,.
\end{equation}
Combining altogether, we thus get
\begin{equation}
G^{\Psi\Psi\,{\cal O}}(X,Y,Z)=C_1\left(1+\frac{[X\cdot\Gamma, Y\cdot\Gamma]}{2X\cdot Y}\right)+C_8\,X\cdot\Gamma\, Z\cdot\Gamma \, Y\cdot\Gamma\,.
\end{equation}
The remaining scalar functions $C_{1,8}$ can be fixed by imposing the scaling transformation \reef{psicon} for Dirac's spinors and $\mathcal{O}(\lambda X)=\lambda^{-\Delta_\mathcal{O}}\,\mathcal{O}(X)$ for the scalar,
\begin{equation}
\label{fermion_fermion_scalar}
G^{\Psi\Psi\,{\cal O}}(X,Y,Z)\equiv\frac{B_1\left(1+\frac{[X\cdot\Gamma, Y\cdot\Gamma]}{2X\cdot Y}\right)+B_2\,\frac{X\cdot\Gamma\, Z\cdot\Gamma \, Y\cdot\Gamma}{\sqrt{X\cdot Y\,X\cdot Z\,Y\cdot Z}}}{(X\cdot Y)
^{\frac{\Delta_{123}-1}{2}}(X\cdot Z)^{\frac{\Delta_{132}}{2}}
(Y\cdot Z)^{\frac{\Delta_{231}}{2}}}\,,
\end{equation}
where $B_{1,2}$ are some constants and $\Delta_{ijk}=\Delta_i+\Delta_j-\Delta_k$ (for $i,j,k=1,2,3$) with $\Delta_{1,2}=\Delta_{\psi_{1,2}},~\Delta_3=\Delta_{\mathcal{O}}$.

Representation of the gamma matrices  \reef{gamma2}  makes it simple to project the above $SO(d,2)$ invariant correlation function onto $\mathbb{R}^{d-1,1}$. For instance, using the relations \reef{relpsi1} and \reef{relpsi2} between the Dirac fields $\psi$ and $\Psi$, we obtain 
\be
 \langle \psi_1(x)\bar\psi_2(y){\cal O}(z)\rangle = (X^+)^{\Delta_1-\frac{1}{2}}(Y^+)^{\Delta_2-\frac{1}{2}}(Z^+)^{\Delta_3}
 \langle \Psi_{1-}(X)\bar\Psi_{2+}(Y){\cal O}(Z)\rangle~.
\ee
Thus we only need to identify $-+$ block of the appropriate supermatrix in \reef{fermion_fermion_scalar}. In particular, up to an overall constant, the term proportional to $B_1$ projects to 
\begin{equation}
G^{\psi\psi\,{\cal O}}_{(1)}(x,y,z)\equiv\frac{\gamma^\mu (x-y)_\mu}{\big((x-y)^2\big)^{\Delta_{123}+1\over 2}
\big((y-z)^2\big)^{\Delta_{231}\over 2}\big((z-x)^2\big)^{\Delta_{132}\over 2}}\,,
\end{equation}
where we used \reef{coor} and 
\begin{equation}
X\cdot Y= -{1\over 2} \, X^+ \, Y^+ \, (x-y)^2~.
\end{equation}

Next let us calculate projection of the term proportional to $B_2$ in (\ref{fermion_fermion_scalar}). It boils down to finding the $-+$ block of 
\begin{equation}
\label{A_d_2_definition}
{\cal A}^{d+2}\equiv (X\cdot\Gamma)\, (Z\cdot\Gamma) \, (Y\cdot\Gamma)
\end{equation}
The only triples of the gamma matrices \reef{gamma2} with non-zero $-+$ blocks are $\Gamma^\mu\Gamma^\nu\Gamma^-$, $\Gamma^-\Gamma^\mu\Gamma^\nu$, $\Gamma^\mu\Gamma^-\Gamma^\nu$ and $\Gamma^-\Gamma^+\Gamma^-$. Hence,
\begin{equation}
{\cal A}^{d+2}_{-+}=X^+Y^+Z^+\Big((x_\mu z_\nu -x_\mu y_\nu+ z_\mu y_\nu)\gamma^\mu \gamma^\nu
-z^2\Big)\,.
\end{equation}
Therefore, up to an overall constant, the $B_2$ term of \reef{fermion_fermion_scalar} projects to
\begin{equation}
\label{Gf_2}
G^{\psi\psi\,{\cal O}}_{(2)}\equiv \frac{(x-z)_\mu(y-z)_\nu \gamma^\mu\gamma^\nu}{\big((x-y)^2\big)^{\Delta_{123}\over 2}\big((x-z)^2\big)^{\Delta_{132}+1\over 2}
\big((y-z)^2\big)^{\Delta_{231}+1\over 2}}\,.
\end{equation}
Combining, we finally obtain
\bea
\label{fermion-fermion-scalar_general}
  &&\langle \psi_1(x)\bar\psi_2(y){\cal O}(z)\rangle = b_1G^{\psi\psi\,{\cal O}}_{(1)}(x,y,z) + b_2 G^{\psi\psi\,{\cal O}}_{(2)} (x,y,z)
  \\
  &&=\frac{1 }{ \big((x-y)^2\big)^{\Delta_{123}\over 2}  \big((y-z)^2\big)^{\Delta_{231}\over2} \big((z-x)^2\big)^{\Delta_{132}\over 2}  }
  \( {b_1\,(\slashed{ x }-\slashed{ y})\over\big((x-y)^2\big)^{1\over 2}}  + {b_2\,(\slashed{ x }-\slashed{ z}) (\slashed{ y }-\slashed{ z})\over \big((x-z)^2(y-z)^2\big)^{1\over 2}}\) 
\nonumber
\eea
where $b_{1,2}$ are some constants and $\slashed x =\gamma^\mu x_\mu$.


\begin{thebibliography}{1234567}

%\cite{Kibble:1976sj}
\bibitem{Kibble:1976sj} 
  T.~W.~B.~Kibble,
  ``Topology of Cosmic Domains and Strings,''
  J.\ Phys.\ A {\bf 9}, 1387 (1976)
  doi:10.1088/0305-4470/9/8/029.
  %%CITATION = doi:10.1088/0305-4470/9/8/029;%%
  %2048 citations counted in INSPIRE as of 26 Mar 2018
  
  %\cite{Zurek:1985qw}
\bibitem{Zurek:1985qw} 
  W.~H.~Zurek,
  ``Cosmological Experiments in Superfluid Helium,''
  Nature {\bf 317}, 505 (1985)
  doi:10.1038/317505a0.
  %%CITATION = doi:10.1038/317505a0;%%
  %330 citations counted in INSPIRE as of 26 Mar 2018

%\cite{Greiner:2002}
\bibitem{Greiner:2002} 
  M.~Greiner, O.~Mandel, T.~W.~H\"ansch, and I.~Bloch,
  ``Collapse and revival of the matter wave field of a Bose--Einstein condensate,''
  Nature {\bf 419}, 51 (2002)
  doi:10.1038/nature00968
  [arXiv:cond-mat/0207196].

  %\cite{Bloch:2007}
\bibitem{Bloch:2007} 
  I.~Bloch, J.~Dalibard, W.~Zwerger,
  ``Many-Body Physics with Ultracold Gases,''
  Rev.\ Mod.\ Phys. {\bf 80}, 885 (2008)
  10.1103/RevModPhys.80.885
   [arXiv:0704.3011 [cond-mat]]. 
   
   %\cite{Polkovnikov:2010}
\bibitem{Polkovnikov:2010} 
  A.~Polkovnikov, K.~Sengupta, A.~Silva, M.~Vengalattore,
  ``Nonequilibrium dynamics of closed interacting quantum systems,''
  Rev.\ Mod.\ Phys. {\bf 83}, 863 (2011)
  10.1103/RevModPhys.83.863
  [arXiv:1007.5331 [cond-mat]].
  
   %\cite{Cazalilla:2011}
\bibitem{Cazalilla:2011} 
  M.~A.~Cazalilla, R.~Citro, T.~Giamarchi, E.~Orignac, M.~Rigol,
  ``One dimensional Bosons: From Condensed Matter Systems to Ultracold Gases,''
  Rev.\ Mod.\ Phys. {\bf 83}, 1405 (2011)
  10.1103/RevModPhys.83.1405
  [arXiv:1101.5337 [cond-mat]].
  
  %\cite{Mitra:2017}
\bibitem{Mitra:2017} 
  A.~Mitra, 
  ``Quantum quench dynamics,''
  [arXiv:1703.09740 [cond-mat]].
  
      %\cite{Srednicki94}
  \bibitem{Srednicki94} 
  M.~Srednicki,
  ``Chaos and quantum thermalization,''
  Phys.\ Rev.\ E.\  {\bf 50}, 888 (1994)
  doi:10.1103/PhysRevE.50.888
  [arXiv:cond-mat/9403051].
  
    %\cite{Fagotti2013}
  \bibitem{Fagotti2013} 
  M.~Fagotti and F.~Essler,
  ``Reduced Density Matrix after a Quantum Quench,''
  Phys.\ Rev.\ B.\  {\bf 87}, 245107 (2013)
  doi:10.1103/PhysRevB.87.245107
  [arXiv:1302.6944 [cond-mat.stat-mech]].

 %\cite{Calabrese:2006rx}
\bibitem{Calabrese:2006rx} 
  P.~Calabrese and J.~L.~Cardy,
  ``Time-dependence of correlation functions following a quantum quench,''
  Phys.\ Rev.\ Lett.\  {\bf 96}, 136801 (2006)
  doi:10.1103/PhysRevLett.96.136801
  [arXiv:cond-mat/0601225].
  %%CITATION = doi:10.1103/PhysRevLett.96.136801;%%
  %242 citations counted in INSPIRE as of 12 Mar 2018 
  
  %\cite{Calabrese:2007rg}
\bibitem{Calabrese:2007rg} 
  P.~Calabrese and J.~Cardy,
  ``Quantum Quenches in Extended Systems,''
  J.\ Stat.\ Mech.\  {\bf 0706}, P06008 (2007)
  doi:10.1088/1742-5468/2007/06/P06008
  [arXiv:0704.1880 [cond-mat.stat-mech]].
  %%CITATION = doi:10.1088/1742-5468/2007/06/P06008;%%
  %128 citations counted in INSPIRE as of 12 Mar 2018
   
%\cite{Delfino:2014qfa}
\bibitem{Delfino:2014qfa} 
  G.~Delfino,
  %``Quantum quenches with integrable pre-quench dynamics,''
  J.\ Phys.\ A {\bf 47}, no. 40, 402001 (2014)
  doi:10.1088/1751-8113/47/40/402001
  [arXiv:1405.6553 [cond-mat.stat-mech]].
  %%CITATION = doi:10.1088/1751-8113/47/40/402001;%%
  %21 citations counted in INSPIRE as of 08 Apr 2018   
   
%\cite{Delfino:2016bln}
\bibitem{Delfino:2016bln} 
  G.~Delfino and J.~Viti,
  %``On the theory of quantum quenches in near-critical systems,''
  J.\ Phys.\ A {\bf 50}, no. 8, 084004 (2017)
  doi:10.1088/1751-8121/aa5660
  [arXiv:1608.07612 [cond-mat.stat-mech]].
  %%CITATION = doi:10.1088/1751-8121/aa5660;%%
  %7 citations counted in INSPIRE as of 08 Apr 2018   
   
   
 %\cite{Chesler:2008hg}
\bibitem{Chesler:2008hg} 
  P.~M.~Chesler and L.~G.~Yaffe,
  ``Horizon formation and far-from-equilibrium isotropization in supersymmetric Yang-Mills plasma,''
  Phys.\ Rev.\ Lett.\  {\bf 102}, 211601 (2009)
  doi:10.1103/PhysRevLett.102.211601
  [arXiv:0812.2053 [hep-th]].
  %%CITATION = doi:10.1103/PhysRevLett.102.211601;%%
  %296 citations counted in INSPIRE as of 12 Mar 2018
  
   %\cite{Basu:2011ft}
\bibitem{Basu:2011ft} 
  P.~Basu and S.~R.~Das,
  ``Quantum Quench across a Holographic Critical Point,''
  JHEP {\bf 1201}, 103 (2012)
  doi:10.1007/JHEP01(2012)103
  [arXiv:1109.3909 [hep-th]].
  %%CITATION = doi:10.1007/JHEP01(2012)103;%%
  %61 citations counted in INSPIRE as of 12 Mar 2018   
   
  %\cite{Buchel:2012gw}
\bibitem{Buchel:2012gw} 
  A.~Buchel, L.~Lehner and R.~C.~Myers,
  ``Thermal quenches in N=2* plasmas,''
  JHEP {\bf 1208}, 049 (2012)
  doi:10.1007/JHEP08(2012)049
  [arXiv:1206.6785 [hep-th]].
  %%CITATION = doi:10.1007/JHEP08(2012)049;%%
  %75 citations counted in INSPIRE as of 12 Mar 2018
  
  %\cite{Buchel:2013lla}
\bibitem{Buchel:2013lla} 
  A.~Buchel, L.~Lehner, R.~C.~Myers and A.~van Niekerk,
  ``Quantum quenches of holographic plasmas,''
  JHEP {\bf 1305}, 067 (2013)
  doi:10.1007/JHEP05(2013)067
  [arXiv:1302.2924 [hep-th]].
  %%CITATION = doi:10.1007/JHEP05(2013)067;%%
  %73 citations counted in INSPIRE as of 12 Mar 2018
  
  %\cite{Buchel:2013gba}
\bibitem{Buchel:2013gba} 
  A.~Buchel, R.~C.~Myers and A.~van Niekerk,
  ``Universality of Abrupt Holographic Quenches,''
  Phys.\ Rev.\ Lett.\  {\bf 111}, 201602 (2013)
  doi:10.1103/PhysRevLett.111.201602
  [arXiv:1307.4740 [hep-th]].
  %%CITATION = doi:10.1103/PhysRevLett.111.201602;%%
  %56 citations counted in INSPIRE as of 12 Mar 2018
  
  %\cite{Das:2014jna}
\bibitem{Das:2014jna} 
  S.~R.~Das, D.~A.~Galante and R.~C.~Myers,
  ``Universal scaling in fast quantum quenches in conformal field theories,''
  Phys.\ Rev.\ Lett.\  {\bf 112}, 171601 (2014)
  doi:10.1103/PhysRevLett.112.171601
  [arXiv:1401.0560 [hep-th]].
  %%CITATION = doi:10.1103/PhysRevLett.112.171601;%%
  %39 citations counted in INSPIRE as of 12 Mar 2018
  
  %\cite{Das:2014hqa}
\bibitem{Das:2014hqa} 
  S.~R.~Das, D.~A.~Galante and R.~C.~Myers,
  ``Universality in fast quantum quenches,''
  JHEP {\bf 1502}, 167 (2015)
  doi:10.1007/JHEP02(2015)167
  [arXiv:1411.7710 [hep-th]].
  %%CITATION = doi:10.1007/JHEP02(2015)167;%%
  %39 citations counted in INSPIRE as of 12 Mar 2018
  
  %\cite{Das:2015jka}
\bibitem{Das:2015jka} 
  S.~R.~Das, D.~A.~Galante and R.~C.~Myers,
  ``Smooth and fast versus instantaneous quenches in quantum field theory,''
  JHEP {\bf 1508}, 073 (2015)
  doi:10.1007/JHEP08(2015)073
  [arXiv:1505.05224 [hep-th]].
  %%CITATION = doi:10.1007/JHEP08(2015)073;%%
  %17 citations counted in INSPIRE as of 12 Mar 2018
 
  %\cite{Das:2016lla}
\bibitem{Das:2016lla} 
  S.~R.~Das, D.~A.~Galante and R.~C.~Myers,
  ``Quantum Quenches in Free Field Theory: Universal Scaling at Any Rate,''
  JHEP {\bf 1605}, 164 (2016)
  doi:10.1007/JHEP05(2016)164
  [arXiv:1602.08547 [hep-th]].
  %%CITATION = doi:10.1007/JHEP05(2016)164;%%
  %8 citations counted in INSPIRE as of 12 Mar 2018
  
%\cite{Das:2016eao}
\bibitem{Das:2016eao} 
  S.~R.~Das,
  ``Old and New Scaling Laws in Quantum Quench,''
  PTEP {\bf 2016}, no. 12, 12C107 (2016)
  doi:10.1093/ptep/ptw146
  [arXiv:1608.04407 [hep-th]].
  %%CITATION = doi:10.1093/ptep/ptw146;%%
  %4 citations counted in INSPIRE as of 26 Mar 2018
  
  %\cite{Das:2017sgp}
\bibitem{Das:2017sgp} 
  D.~Das, S.~R.~Das, D.~A.~Galante, R.~C.~Myers and K.~Sengupta,
  ``An exactly solvable quench protocol for integrable spin models,''
  JHEP {\bf 1711}, 157 (2017)
  doi:10.1007/JHEP11(2017)157
  [arXiv:1706.02322 [hep-th]].
  %%CITATION = doi:10.1007/JHEP11(2017)157;%%
  %3 citations counted in INSPIRE as of 12 Mar 2018
  
  %\cite{Dymarsky:2017awt}
\bibitem{Dymarsky:2017awt} 
  A.~Dymarsky and M.~Smolkin,
  ``Universality of fast quenches from the conformal perturbation theory,''
  JHEP {\bf 1801}, 112 (2018)
  doi:10.1007/JHEP01(2018)112
  [arXiv:1709.08654 [hep-th]].
  %%CITATION = doi:10.1007/JHEP01(2018)112;%%
  

%\cite{Weinberg:2010fx}
\bibitem{Weinberg:2010fx} 
  S.~Weinberg,
  %``Six-dimensional Methods for Four-dimensional Conformal Field Theories,''
  Phys.\ Rev.\ D {\bf 82}, 045031 (2010)
  doi:10.1103/PhysRevD.82.045031
  [arXiv:1006.3480 [hep-th]].
  %%CITATION = doi:10.1103/PhysRevD.82.045031;%%
  %110 citations counted in INSPIRE as of 27 Nov 2017

%\cite{Costa:2011mg}
\bibitem{Costa:2011mg} 
  M.~S.~Costa, J.~Penedones, D.~Poland and S.~Rychkov,
  ``Spinning Conformal Correlators,''
  JHEP {\bf 1111}, 071 (2011)
  doi:10.1007/JHEP11(2011)071
  [arXiv:1107.3554 [hep-th]].
  %%CITATION = doi:10.1007/JHEP11(2011)071;%%
  %172 citations counted in INSPIRE as of 13 Sep 2017


%\cite{Rychkov:2016iqz}
\bibitem{Rychkov:2016iqz} 
  S.~Rychkov,
  ``EPFL Lectures on Conformal Field Theory in D$\geq$ 3 Dimensions,''
  doi:10.1007/978-3-319-43626-5
  arXiv:1601.05000 [hep-th].
  %%CITATION = doi:10.1007/978-3-319-43626-5;%%
  %66 citations counted in INSPIRE as of 13 Sep 2017
  

  


  
  %\cite{Isono:2017grm}
\bibitem{Isono:2017grm} 
  H.~Isono,
  ``On conformal correlators and blocks with spinors in general dimensions,''
  Phys.\ Rev.\ D {\bf 96}, no. 6, 065011 (2017)
  doi:10.1103/PhysRevD.96.065011
  [arXiv:1706.02835 [hep-th]].
  %%CITATION = doi:10.1103/PhysRevD.96.065011;%%
  

\end{thebibliography}
\end{document}